\documentclass[twocolumn,floatfix,prl]{revtex4-2}%
\usepackage[dvipdfmx]{graphicx}%
\usepackage{amsmath}%
\usepackage{amsfonts}%
\usepackage{amssymb}
\usepackage{bm}
\usepackage{color}
\usepackage{ulem}

\def\s{{\sigma}}
\def\e{{\epsilon}}
\def\k{{ {\bm k} }}
\def\p{{ {\bm p} }}
\def\q{{ {\bm q} }}
\def\Q{{ {\bm Q} }}

\def\0{{ {\bm 0} }}
\def\w{{\omega}}
\def\a{{\alpha}}
\def\b{{\beta}}

\def\r{{ {\bm r} }}

\allowdisplaybreaks[4]

\begin{document}
\title{
\color{black}
$SU(4)$ \color{black}Valley + Spin Fluctuation Interference
Mechanism for Nematic Order in \color{black}Magic Angle \color{black}Twisted Bilayer Graphene: Impact of Vertex Corrections \color{black}}
\author{
Seiichiro Onari and Hiroshi Kontani
}

\date{\today }

\begin{abstract}
In the magic angle twisted bilayer graphene (MATBG), one of the most
remarkable observations is the $C_3$-symmetry-breaking nematic state.
\color{black}We identify that the nematicity in MATBG is the $E$-symmetry
 ferro bond order, which is the modulation of correlated hopping
 integrals owing to the $E$-symmetry particle-hole pairing condensation
\color{black}
%They originate from the spontaneous symmetry breaking in the
% non-local self-energy. 
% driven by the electron correlation. 
The nematicity in MATBG originates from  
prominent quantum interference among 
\color{black} $SU(4)$ valley+spin composite fluctuations. \color{black}
This novel ``valley + spin fluctuation interference mechanism'' is
 revealed by the density wave equation analysis for realistic
 multiorbital Hubbard model for MATBG. \color{black}We find that the nematic
 state is robust once three van Hove singularity points exist in each
 valley. \color{black}
This interference mechanism also causes novel time-reversal-symmetry-broken
valley polarization accompanied by a charge loop current. 
We discuss interesting similarities and differences 
between MATBG and Fe-based superconductors.
\end{abstract}

\address{
 Department of Physics, Nagoya University,
Furo-cho, Nagoya 464-8602, Japan. 
}
 
%\pacs{74.70.Xa, 75.25.Dk, 74.20.Pq} 
%74.20.Pq Electronic structure calculations

\sloppy

\maketitle

%%%%%%%%%%%%%%%%%%
%Introduction
%%%%%%%%%%%%%%%%%%
%nematic electronic state
The emergence of the exotic electronic states in the
magic angle ($\sim 1.1^\circ$) twisted bilayer graphene (MATBG) opens a novel platform of 
strongly correlated electron systems \cite{Cao1,Cao2,Yankowitz,Lu}.
Since the moir\'{e} pattern in MATBG makes superlattice, nearly flat
band due to the multi band folding appears around the charge neutrality.
The nearly flat band provides the strong correlation system with many van
Hove singularity (VHS) points.
% $U\gtrsim W$ even if the Coulomb interaction is weak. 
The superconducting phase broadly appears 
near the VHS filling $n\sim\pm 2$, where $n$ denotes number of
electrons in the moir\'{e} superlattice unit cell, and $n=0$ corresponds to the
charge neutrality. 
A lot of important theoretical studies have been performed in the last
few years
\cite{Xu,Fernandes-super,Isobe-super,Dodaro,Kang,Seo,Roy,Ray,You,Zhang,Chubukov-super,Fernandes-super2,Lin,MacDonald,Bultinck}.

Recently, the ferro $(\q=\0)$ $C_3$-symmetry-breaking nematic state
has been observed by STM and resistivity anisotropy measurements in MATBG \cite{nematic-TBG,nematic-TBG2,nematic-TBG3,nematic-TBG4}.
In the vicinity of the VHS filling, the
electronic nematic state appears in the metallic phase
\cite{nematic-TBG,nematic-TBG4}. 
To explain the nematicity in MATBG, the acoustic phonon
mechanism\cite{Fernandes-TBG} have been proposed by restricting to the ferro order $(\q=\0)$.
\color{black}Also, the electron correlation mechanism has been studied using the mean
field theory \cite{Fernandes-TBG2}. It is well-known that the instability in 
 the mean-field theory [= the random-phase-approximation (RPA)] occurs at the
 nesting vector $\q\ne\bm{0}$. Thus, the $\q=\bm{0}$ nematic order
 requires beyond the RPA. 
The following fundamental questions remain open problems: 
What types of electron correlations drive the nematicity? \color{black}
Why the nematic order is selected over rich degrees of freedom in MATBG? 
 \color{black}These questions on the nematicity of MATBG have been still open even after the 
pioneering beyond-RPA analyses using the renormalization group (RG) methods \cite{Isobe-super,Chubukov-nematic}.\color{black} 

%Also, why the nematic order is $\q=\0$? 
%\color{black} Although nematicity in ``untwisted bilayer graphene'' has been
%discussed \cite{Nem-graphene1,Nem-graphene2,Nem-graphene3,Nem-graphene4},
%its microscopic driving force is still unclear. \color{black}

\begin{figure}[!htb]
\includegraphics[width=.99\linewidth]{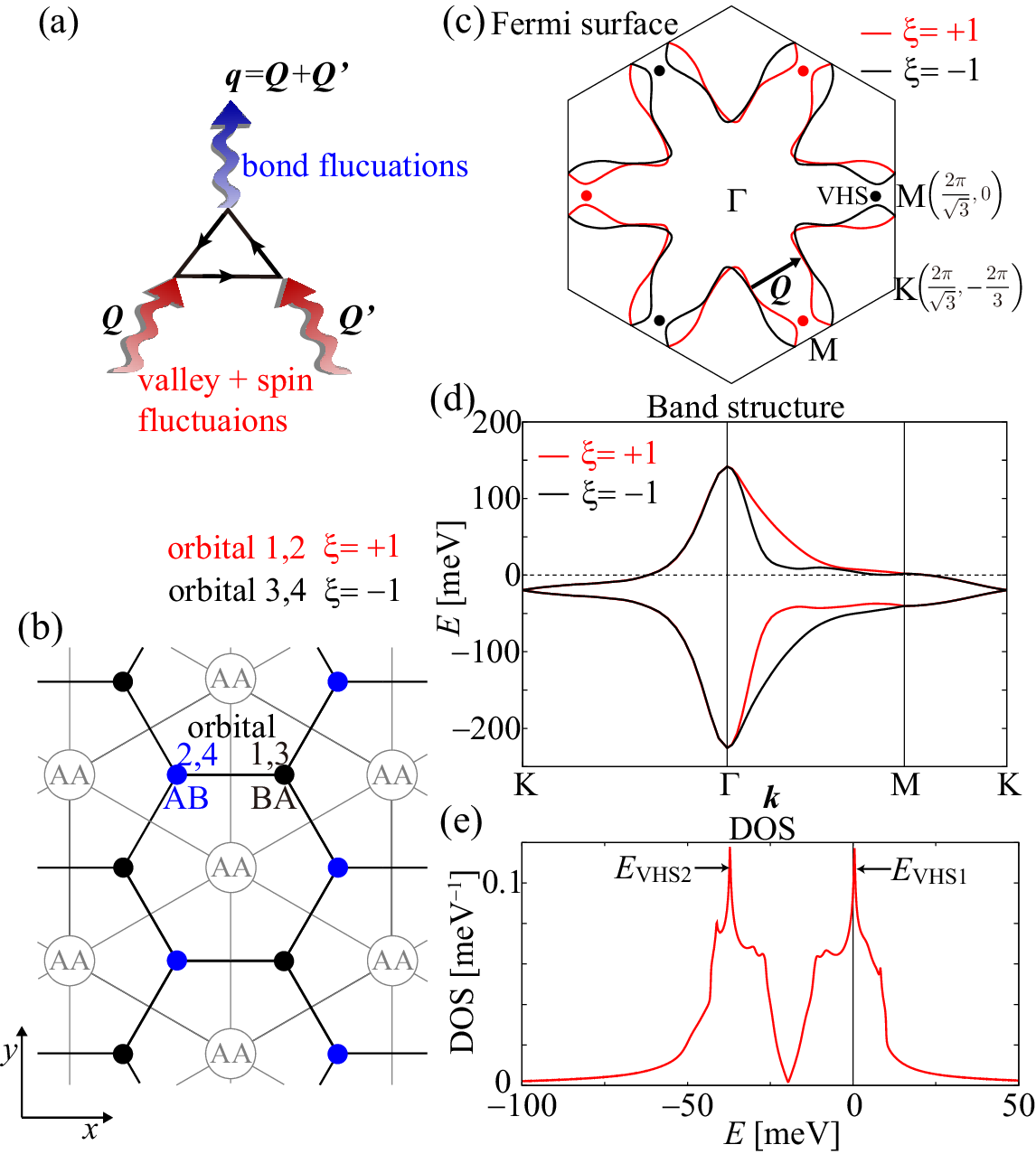}
\caption{
%(color online)
(a) Quantum interference between valley+spin fluctuations with the wavevector $\Q$ and
 $\Q'$, which induces the bond fluctuations with
 $\bm{q}=\Q+\Q'$.
(b) Moir\'{e} superlattice spanned by the AA spots. Wannier orbitals 1, 3 and
 2, 4 are centered at the BA (black dots) and AB (blue dots) spots, respectively.
 (c) FSs of MATBG for $n=2.0$, where red (black) lines and 
 dots denote the valley and 
 the VHS points for $\xi=+1$ $(-1)$, respectively. 
The vector $\bm{Q}$ is the nesting vector. (d) Band structure and 
 (e) DOS for $n=2.0$, which has peaks at the two VHS energies $E_{\rm VHS1}$ and $E_{\rm VHS2}$.
}
\label{fig:FS}
\end{figure}

The nematic orders are also realized in Fe-based and cuprate superconductors
\cite{Johnston,Mizuguchi,Cu-nematic}. \color{black}
The intertwined-order\cite{Intertwined,Davis-DHLee}, spin-nematic/vestigial-order \color{black} 
\cite{Fernandes,Fernandes-122,DHLee,QSi,Valenti,Fang,Fernandes-review}, and 
orbital/bond-order \cite{Kruger,PP,WKu,Kontani-PRL,Onari-SCVC,Onari-form,Yamakawa-PRX,Onari-B2g,JP,Fanfarillo,Chubukov-FeSe,Chubukov-RG,Tsuchiizu-Cu}
scenarios have been applied to solve this issue.
In the latter scenario, the nematic orbital/bond orders
are generated by the paramagnon interference shown in Fig. \ref{fig:FS}(a)
\cite{Onari-SCVC,Onari-form,Yamakawa-PRX,Onari-B2g,Tsuchiizu-Cu}. 
%The bond order
%is the modulation of correlated hopping induced by the many body effect. The interference is expressed
%by the Aslamazov--Larkin (AL) vertex correction (VC) due to spin
%fluctuations, and its 
Its significance has been confirmed by the functional RG studies 
\cite{Chubukov-FeSe,Chubukov-RG,Tsuchiizu-Cu,Tsuchiizu-Ru1}.
This mechanism may also be applicable to MATBG.
On the other hand, MATBG has two significant characteristics 
distinct from usual transition metal compounds;
(i) presence of the valley degree of freedom $\xi$, and 
(ii) absence of on-site Hund's coupling $J=0$ 
\color{black}
\cite{Koshino,Markus}.
\color{black}
%As for (i), the Wannier orbitals 1 and 2 (3 and 4) in
%Fig. \ref{fig:FS}(b) are labeled as the valley $\xi=+1$ $(-1)$. The FSs
%are also labeled by the valley $\xi$ since inter-valley hopping integrals are a%bsent. The valley $\xi$ changes its sign under the time reversal operation.
%As for (ii), the intra- and inter-valley on-site Coulomb repulsions are exactly
%the same $(U=U')$, and the Hund's coupling is zero \cite{Koshino,Markus}. 
By focusing on both (i) and (ii), 
we explain why rich unconventional density waves appear in MATBG.

In this paper, we study the origin of the nematic state in MATBG based
on microscopic analysis.
% based on the interference mechanism. 
Thanks to the two significant
features, (i) the presence of the valley and (ii) the absence of $J$,
it is driven by interferences among \color{black}$SU(4)$ valley+spin composite fluctuations. \color{black}
% due to the Hund's less nature of the MATBG. 
This interference mechanism also causes
the time-reversal-symmetry-broken valley polarization 
%near the nematic phase.
accompanied by a novel charge loop current.
% that can be measured by several experimental techniques.
This study reveals similarities and differences 
between MATBG and Fe-based superconductors.

%%%%%%%%%%%%%%%%%%%%%%%%%

As for the character (i), the Wannier orbitals 1 and 2 (3 and 4) in
Fig. \ref{fig:FS}(b) are labeled as the valley $\xi=+1$ $(-1)$. The
Fermi surfaces (FSs)
are also labeled by the valley $\xi$ since inter-valley hopping integrals are absent. The valley $\xi$ changes its sign under the time reversal operation.
As for the character (ii), 
the intra- and inter-valley on-site Coulomb repulsions are exactly
the same $(U=U')$ and the Hund's coupling is zero.
%\color{black}Since the wave function of $\xi=+1$ is the complex conjugate
%of that of $\xi=-1$, density of $\xi=+1$ is identical to that of
%$\xi=-1$, and inter-valley $U'$ is the same as intra-valley $U$.
%Inter-valley $J=0$ is approximately given by the strong oscillation of phase in the
%integral of the Coulomb interaction.
%\cite{Koshino,Markus}
%\color{black}
Both (i) and (ii) are key ingredients in the rich
unconventional density waves obtained in the present study.

%%%%%%%%%%%%%%%%%%%%%%%%%

%%%%%%%%%%%%%%%%%%%%%%%%%%%
% Model and Hamiltonian
%%%%%%%%%%%%%%%%%%%%%%%%%%%
We analyze the following two-dimensional 
four-orbital Hubbard model \cite{Onari-form}:
\begin{eqnarray}
H=H^0+H', \ \ \
\label{eqn:Ham}
\end{eqnarray}
where 
$H^0$ is the first-principles model for MATBG in
Ref. \cite{Koshino} with minimum additional terms to make $n_{\rm
VHS}\sim2$, as we explain in Supplementary
Material (SM) A \cite{SM}. Figure \ref{fig:FS}(b) shows Moir\'{e}
superlattice spanned by the AA spots. We define the distance between the
nearest AA
spots as 1. At the AB (BA) spots, the A
(B) sublattice in upper layer just locates above B (A)
sublattice in lower one. 
The centers
of Wannier orbitals $1$, $3$ and $2$, $4$ locate at the BA and AB spots, respectively.
The orbitals $1$ and $2$ $(\xi=+1)$ are transformed to
the orbitals $3$ and $4$ $(\xi=-1)$ by the time-reversal
operation, respectively.
Each valley
is independent in $H^0$: $H^0=H^0_{\xi=+1}+H^0_{\xi=-1}$.
$H'$ is the Coulomb interaction. We consider the intra-valley local Coulomb
interaction $U$ and inter-valley one $U'$ on the same site.
The relation
$U'=U$ $(J=0)$ is satisfied in the Wannier orbitals of MATBG \cite{Koshino,Markus}.
Details of model and interaction are presented in SM A \cite{SM}.

%%%%%%%%%%%%%%%%%%%%%%%%%%%%%%%%%
%%%%%%%%%%%%%%%%%%%%%%%%%%%%%%%%%
Here, we study the electronic states at $n=2.0$, which is close to the
VHS filling $(n_{\rm VHS}=2.07)$. Figure \ref{fig:FS}(c) shows the FSs of MATBG for $\xi=\pm 1$. The weights of two same-valley orbitals on each FS are
almost the same. For each valley, there are three VHS points
at $E_{\rm VHS1}\sim 0.5$ meV, which locate near the FS around the M points, as shown in Fig. \ref{fig:FS}(c).
Figures \ref{fig:FS}(d) and (e) show the band structure with $\xi=\pm1$ and
the total DOS, respectively. \color{black}Split of the two VHS energies $E_{\rm VHS1}-E_{\rm VHS2}\sim 50$meV corresponds to the effective
bandwidth, which is consistent with the STM measurement \cite{nematic-TBG}.
\color{black}

We calculate the spin (charge) susceptibilities ${\hat \chi}^{s(c)}(q)$
for $q=(\q,\w_m=2m\pi T)$ based on the RPA.
Details of formulation are presented in SM A \cite{SM}.
${\hat
\chi}^{s(c)}(q)\propto(1-\a_{s(c)})^{-1}$ is given by the spin (charge)
Stoner factor $\a_{s(c)}$. $\a_{s(c)}=1$ corresponds to spin (charge)-ordered
state.
In the present study, $\a_s=\a_c=\a$ is satisfied due to the relation
$U'=U$ $(J=0)$ \cite{Kontani-PRL}.
Hereafter, we set $T=1.5$ meV. We fix $\a=0.83$, which corresponds to
moderately correlated region, by
setting the solo model parameter $U=39$
$(42)$ meV for $n=2.0$ $(2.4)$.
%$\a=0.83$ corresponds to the moderately correlated region.

\begin{figure}[!htb]
\includegraphics[width=.99\linewidth]{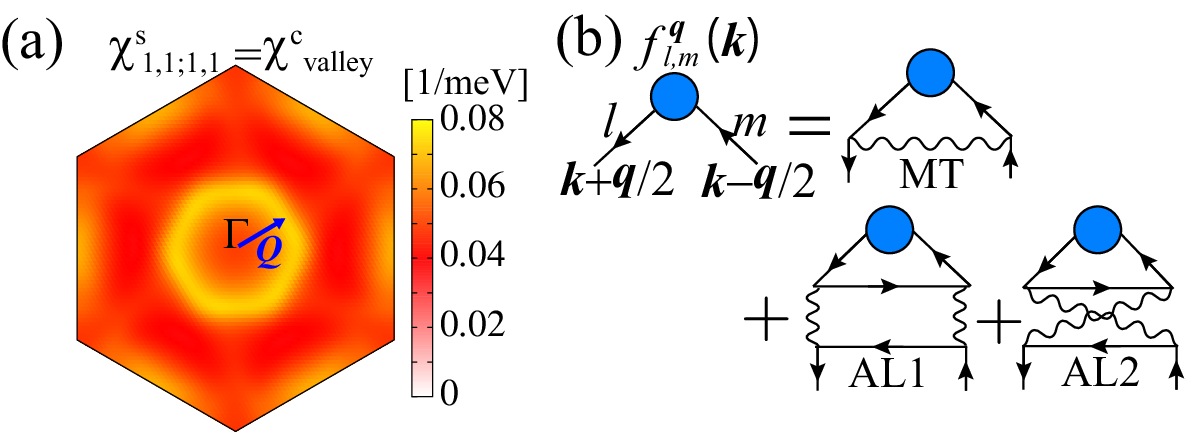}
 \caption{
(a) $\q$ dependences of 
 $\chi^{s}_{1,1;1,1}(\q,0)=\chi^c_{\rm
valley}(\q,0)$ given by the RPA for $n=2.0$. 
(b) Feynman diagrams of the DW equation. Each wavy line represents valley+spin
 fluctuation-mediated interaction.
}
\label{fig:chi}
\end{figure}

Figure \ref{fig:chi}(a) shows the obtained spin susceptibility
$\chi^{s}_{1,1;1,1}(\q,0)$, which shows broad maximum peak at the
intra-orbital nesting $\Q$ around the VHS points.
%KON
We stress that the valley susceptibility $\chi^c_{\rm valley}
\equiv\chi^{c}_{1,1;1,1}-\chi^{c}_{1,1;3,3}$ is exactly the
same as $\chi^{s}_{1,1;1,1}$, as the consequence of the 
(approximate) $SU(4)$ symmetry of the MATBG model, 
as we explain in SM B \cite{SM}.

%KON
%This result is explained by the mean field energy for the valley channel $-\frac{2U'-U}{8}\langle
%n_{1\uparrow}+n_{1\downarrow}-n_{3\uparrow}-n_{3\downarrow}\rangle$ and
%that for the
% spin channel $-\frac{U}{8}\langle
%n_{1\uparrow}-n_{1\downarrow}+n_{3\uparrow}-n_{3\downarrow}\rangle$,
%both of which have
%the same interaction $-\frac{U}{8}$. Here,
% $n_{l\sigma}$ denotes number of electrons for the orbital $l$ and
%spin $\sigma$.
%The absence of the Hund's coupling in TBG is
%significantly different from usual transition metal compounds $(J/U\gtrsim 0.1)$.

%%%%%%%%%%%%%%%%%%%%%%%%%%%%%%%%%
%%%%%%%%%%%%%%%%%%%%%%%%%%%%%%%%%

Hereafter, we derive the most strongest charge-channel density-wave (DW)
instability, without assuming the order parameter and its wave vector.
For this purpose, we use the DW equation method developed in Refs. \cite{Onari-form,Onari-B2g,Kontani-sLC}. 
The linearized DW equation is given as
%self-consistent equations with respect to the symmetry-breaking
%self-energy $\Delta\hat{\Sigma}^{\q}$, and obtain the linearized CDW equation:
%
\begin{eqnarray}
  \!\!\!\!\lambda_\q f^\q_{l,l'}(k)&=& \frac{T}{N}
 \!\!\sum_{k',m,m'} {K}^{c,\bm{q}}_{l,l';m,m'}(k,k')f^\q_{m,m'}(k'),
 \label{eqn:linearized}
\end{eqnarray}
\color{black}
where the kernel function is 
$K^{c,\bm{q}}_{l,l';m,m'}(k,k')=-\sum_{m_1,m_2}\!
I^{c,\bm{q}}_{l,l';m_1,m_2}(k,k')G_{m_1,m}\left(k'+\frac{\bm{q}}{2}\right)\!G_{m',m_2}(k'-\frac{\q}{2})$. $\hat{I}^{c,\bm{q}}(k,k')$
is the charge-channel irreducible interaction. 
$\lambda_\q$ is the eigenvalue of the form factor $\hat{f}^\q$,
which represents the symmetry breaking in the self-energy,
or equivalently, symmetry-breaking particle-hole pairing condensation. 
To satisfy the conservation laws \cite{Luttinger,Baym}, one has to set $\hat{I}^{c,\q=\bm{0}}(k,k')={\delta^2\Phi}/{\delta
 \hat{G}(k)\delta \hat{G}(k')}$,
where $\Phi$ is the Luttinger-Ward function. Here, we apply the one-loop
approximation for $\Phi$, and the derived $\hat{I}^{c,\bm{q}}(k,k')$ is given in SM A \cite{SM}. \color{black}
%Feynman diagram of the linearized CDW equation (\ref{eqn:linearized}) is depicted in Fig. \ref{fig:FS} (e).
The charge-channel DW with wavevector $\q$ is established when
the largest $\lambda_\q=1$.
The charge-channel DW susceptibility is proportional to
 $(1-\lambda_\q)^{-1}$. Therefore, $\lambda_\q$ represents the strength
 of the DW instability.
The Maki--Thompson (MT) terms and Aslamazov--Larkin (AL) terms shown in
Fig. \ref{fig:chi}(b) are included in the kernel function.
\color{black}The AL term is magnified by the convolution of spin/charge susceptibilities. 
In a simple single-orbital model, for example, the charge- and
spin-channel AL terms are proportional to
$C^c(q)\sim 3C^{s,s}(q)+C^{c,c}(q)$ and $C^s(q)\sim 2C^{s,c}(q)$, respectively,
where $C^{a,b} (q) = \sum_p \chi^a(p) \chi^{b}(-p+q)$ and $a,b=c$ or $s$.
Apparently, $C^{c(s)}(q)$ takes the largest value at $\q=\bm{0}$. (i.e.,
$\Q=-\Q'$ in Fig. \ref{fig:FS}(a).) \color{black}
%
%In the presence of valley and spin fluctuations at $\q=\Q$ and $\Q'$,
%the AL terms composed of two wavy lines are strongly enhanced in proportion to
%$\sum_p\chi(p)\chi(-p+q_c)$, where $\chi(q)$ is the valley or spin
%susceptibility. 
Therefore, the quantum interference mechanism
by the AL terms
 causes novel DW orders at $\q=\bm{0}$ in various Hubbard models
 \cite{Onari-SCVC,Onari-form,Yamakawa-PRX,Onari-B2g,Tsuchiizu-Cu}, and this mechanism will be significant for MATBG.

\begin{figure}[!htb]
\includegraphics[width=.99\linewidth]{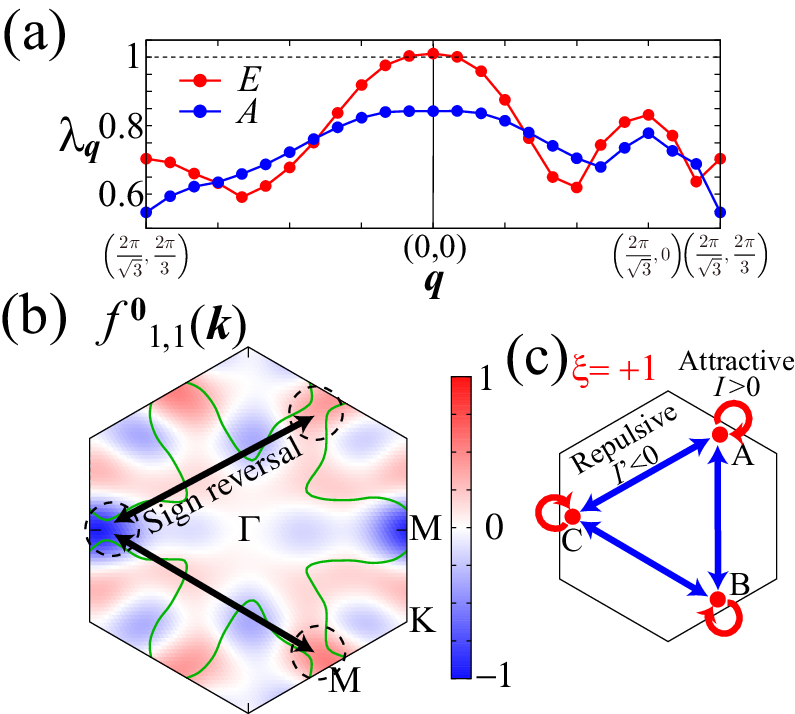}
 \caption{\color{black}
(a) Obtained $\q$ dependences of $\lambda_{\q}$ for the $E$ and $A$
 symmetries for $n=2.0$. Maximum peak of $E$-symmetry at $\q=\0$ means the emergence of the ferro nematic order.
(b) Dominant form factor $f_{1,1}^{\bm{0}}(\k)$ in the Brillouin zone.
The green lines indicate FSs for $\xi=+1$. The intra- (inter-) valley
 relation $f^{\bm{0}}_{1,1}(\k)=f^{\bm{0}}_{2,2}(\k)$
 $(f^{\bm{0}}_{1,1}(\k)=f^{\bm{0}}_{3,3}(-\k))$ is satisfied.
Black arrows show the sign-reversal between the VHS points.
(c) Schematic picture of intra-VHS attractive $I_{\rm intra}>0$ and inter-VHS
 repulsive $I_{\rm inter}<0$ interactions for orbital $1$ in the simple three VHS model.
\color{black}}
\label{fig:Sigma}
\end{figure}

%From the solution ${\hat f}^\q(k)$ of the DW equation, we
%derive the static form factor ${\hat f}^\q(\k)$ at $\e=0$ based on 
%analytic continuation.
Figure \ref{fig:Sigma}(a) shows the $\q$ dependences of the obtained
$\lambda_\q$ for the $C_3$ symmetry breaking $E$-symmetry and $C_3$
$A$-symmetry. \color{black}
Importantly, the ferro-nematic order is realized
$(\lambda_{\bm{0}}=1)$ even when
the SDW/CDW susceptibilities are small $(\alpha \ll 1)$ 
in the present theory. 
\color{black}
Figure \ref{fig:Sigma}(b) shows
the dominant static form factor $f^{\bm{0}}_{1,1}(\k)$, which is derived from the analytic
continuation of $\hat{f}^{\q}(k)$. 
The obtained form factor has no inter-valley component, and satisfies the time-reversal
invariance.
% $[f^{\bm{0}}_{1,1}(\k)=f^{\bm{0}}_{3,3}(-\k)]$ and $f^{\bm{0}}_{1,2}(\k)=f^{\bm{0}*}_{3,4}(-\k)$. 
The obtained
$\hat{f}^{\bm{0}}(\k)$ belongs to the two-dimensional $E$ representation, and its partner $\hat{f}'^{\bm{0}}(\k)$
is shown in SM C \cite{SM}. 
Thus, the direction of nematicity can be rotated by making the linear
combination of
$\hat{f}^{\bm{0}}(\k)$ and $\hat{f}'^{\bm{0}}(\k)$, it will be fixed by the anharmonic phonons and/or the fourth-order terms in the Ginzburg-Landau free energy.

The obtained nematic state is mainly even parity as recognized in
Fig. \ref{fig:Sigma}(b). However, sizable odd-parity component is mixed
 due to absence of the inversion symmetry. To show this, we examine the
 form factors in real space $\tilde{f}_{l,m}(\r)=\frac{1}{N}\sum_{\k}
f^{\bm{0}}_{l,m}(\k)e^{i\k\cdot\r}$ shown in
Fig. S3 in SM C \cite{SM}. 
Its real part Re$\tilde{f}_{1,1}(\r)$ has even parity,
which gives the bond order ($=$ modulation of the correlated hopping integrals). On the other
hand, its imaginary part Im$\tilde{f}_{1,1}(\r)$ has odd parity, which
gives the valley current due to the time-reversal invariance \cite{Kontani-sLC,Tazai}. In the valley current state, the charge
current in one valley is
canceled by that in the opposite valley. \color{black}

The $\q=\0$ nematic order originates from the AL type quantum
interference between the valley+spin fluctuations with $\Q\approx -\Q'$
 shown in Fig. \ref{fig:FS}(a). 
%KON
\color{black}
As we explain in SM B \cite{SM},
the fifteen-channel valley+spin susceptibilities 
$\chi_l(\q)$ ($l=1\sim15$) are equally enhanced,
by reflecting the approximate $SU(4)$ symmetry of the model.
Here, $l=1\sim3$ for valley, $l=4\sim6$ for spin,
$l=7\sim15$ for valley+spin composite channels.
The interference among these $SU(4)$ valley+spin fluctuations \cite{Tazai-CeB6}
causes strong nematic criticality efficiently.  \color{black}
In contrast, only spin fluctuations contribute in
Fe-based superconductors with $J\gtrsim 0.1$
\cite{Onari-SCVC,Onari-form,Yamakawa-PRX,Onari-B2g,JP,Fanfarillo,Chubukov-FeSe,Chubukov-RG}
and cuprate superconductors with single orbital \cite{Tsuchiizu-Cu}.
For this reason, the nematic order is more easily realized in MATBG.

\color{black}The mechanism of the $E$-symmetry nematic
order is clearly explained by focusing on the three VHS points on each
valley as follows. The importance of the VHS in MATBG has been
clarified in the RG studies \cite{Isobe-super,Chubukov-nematic}.
Figure \ref{fig:Sigma}(c) shows the intra-VHS attractive interaction $I>0$ 
and inter-VHS repulsive interaction $I'<0$ driven mainly by the AL
terms, as we describe in SM D 1 \cite{SM}. 
When the interaction is restricted for orbital 1, 
%kernel in the DW Eq. (\ref{eqn:linearized}) is given by $3\times 3$
%matrix with $g^{\bm{0}}_{1,1;1,1}<0$ in the three VHS model,
 solved eigenvalues in the simple three VHS model are
doubly degenerate $\lambda_E \propto I-I'$ and non-degenerate
$\lambda_A \propto I+2I'$. The former $\lambda_E$
has degenerate form factors
$(f_A,f_B,f_C)\propto(1,1,-2)/\sqrt{6}$, $(1,-1,0)/\sqrt{2}$, where $f_X$ denotes the form factor
at the VHS $X(=A,B,C)$ point.
These form factors correspond to $E$-symmetry $f^{\bm{0}}_{1,1}$ in Fig. \ref{fig:Sigma}(b) and
$f'^{\bm{0}}_{1,1}$ in SM C \cite{SM}.
The latter $\lambda_A$ has $(f_A,f_B,f_C)\propto(1,1,1)/\sqrt{3}$,
which corresponds to the  $A$-symmetry. 
Thus, dominant nematic $E$-symmetry with $\lambda_E(>\lambda_A)$ is explained by
the relations $I>0$ and $I'<0$ derived by the valley+spin
interference. This nematic state is robust independently of the shape of
FSs once three VHS points exist in each valley.
 \color{black}

%%%%%%%%%%%%%%%%%%%%%%%%%%%%%%%%%
%%%%%%%%%%%%%%%%%%%%%%%%%%%%%%%%%

One of the main merit of the present bond-order theory is that the
ferro ($\q=\0$) order is naturally obtained. Moreover, the present
bond-order theory can cooperate with the
phonon mechanism proposed in Ref. \cite{Fernandes-TBG}, as discussed in Ref. \cite{Kontani-Phonon}, while the phonon mechanism alone may give the bond order at $\q\approx\Q$.
The nematic state was also discussed from the side of electron
correlation by using the RG theory \color{black}focusing on the VHS points
\cite{Chubukov-nematic}. The AL-type vertex correction (VC) are
included in this theory \cite{Chubukov-FeSe,Chubukov-RG}. \color{black}
Therefore, the difference between the results of the present theory
and  those in Ref. \cite{Chubukov-nematic} may originate from the difference of the theoretical models.

\color{black}
The present interference mechanism predicts that the nematic order in
MATBG is established under very weak spin fluctuations, which is
reminiscent of the nematicity without magnetism in FeSe
\cite{Onari-form,Yamakawa-PRX}. This result is very different
from the vestigial nematic scenario under sufficiently long spin
correlation length \cite{Fernandes}. 

Here, we explain the robustness of the obtained nematic bond order,
which is uniquely obtained near the VHS filling $(-0.3\lesssim n-n_{\rm
VHS}\lesssim 0)$ in the present MATBG model.
The nematic bond order also uniquely appears in the original MATBG model
for $n\sim n_{\rm VHS}=0.7$ \cite{Koshino}, as we demonstrate in SM D 2 \cite{SM}.
Thus, the nematic bond order is robust in the case of $n\sim n_{\rm
VHS}$, insensitively to the FS topology and position of the
VHS points. This robustness is consistent with the intuitive explanation
in Fig. \ref{fig:Sigma}(c) by focusing on the VHS points.
Furthermore, the nematicity is stabilized by the relation $U=U'$ $(J\ll
U)$ in MATBG due to the contribution of the valley fluctuation
interference; see SM E \cite{SM}.

\begin{figure}[ht]
\includegraphics[width=.99\linewidth]{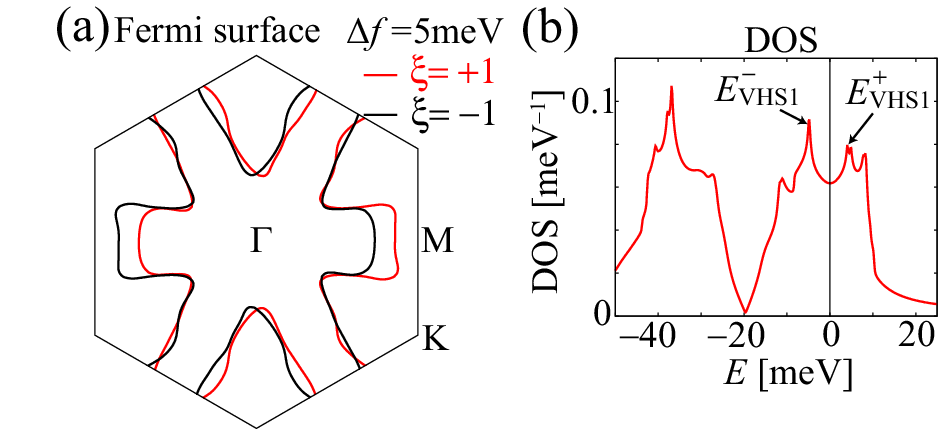}
\caption{
(a) FSs under the nematic order $\Delta
 f=5$ meV, where red (black) lines denote $\xi=+1(-1)$.
(b) DOS for $\Delta f=5$ meV, where the VHS energy $E_{\rm VHS1}$ splits into $E_{\rm VHS1}^\pm\sim \pm\Delta f$.
}
\label{fig:nematic}
\end{figure}

In the following, we explain the nematic ordered state.
We denote $\Delta f$ as the maximum value of $|\hat{f}^{\bm{0}}(\k)|$.
Figure \ref{fig:nematic} (a) shows FSs under the nematic order.
We confirm that the $C_3$ symmetry is broken, and
strong anisotropy appears along the $x$ axis, which is consistent with
experiments \cite{nematic-TBG,nematic-TBG2,nematic-TBG3,nematic-TBG4}.
The DOS under the nematic order $\Delta f=5$ meV is shown in
Fig. \ref{fig:nematic} (b). The energy of the VHS splits into $E_{\rm VHS1}^\pm\sim\pm\Delta
f$ due to the $\k$ dependence of
$\hat{f}^{\bm{0}}(\k)$. The dip structure in the DOS near the
Fermi energy $E=0$ is
consistent with STM measurement \cite{nematic-TBG}.

%Figure \ref{fig:nematic} (c) shows the current from
%$\0$ to $\r$ for orbital 1 in the unit of $1$meV/$\hbar$:
%$j_{1,1}(\r)=-2{\rm Im}\langle
%[\tilde{h}^0_{1,1}(\r)+\tilde{f}_{1,1}(\r)]c^\dagger_1(\r)c_1(\0)
%\rangle$  for $\Delta
%f=5$meV, where $\tilde{h}^0_{1,1}(\r)$ is the hopping.  We confirm that
%the nematic $j_{1,1}(\r)$ breaks microscopically $C_3$
%symmetry as shown in schematic picture of dominant $j_{1,1}(\r)$ in
%Fig. \ref{fig:nematic} (d). 
%The current is canceled by the time reversal current
%$j_{3,3}(\r)$, while the microscopic valley current $j_{1,1}(\r)-j_{3,3}(\r)$
%remains.
%However, macroscopic net current for the orbitals $1$ and $2$ is absent even in the single valley $\xi=+1$, 
%since the obtained current is the nematic loop current as shown in
%Fig. \ref{fig:nematic} (d).

%%%%%%%%%%%%%%%%%%%%%%%%%%%%%%%%%
%%%%%%%%%%%%%%%%%%%%%%%%%%%%%%%%%

\begin{figure}[!htb]
\includegraphics[width=.9\linewidth]{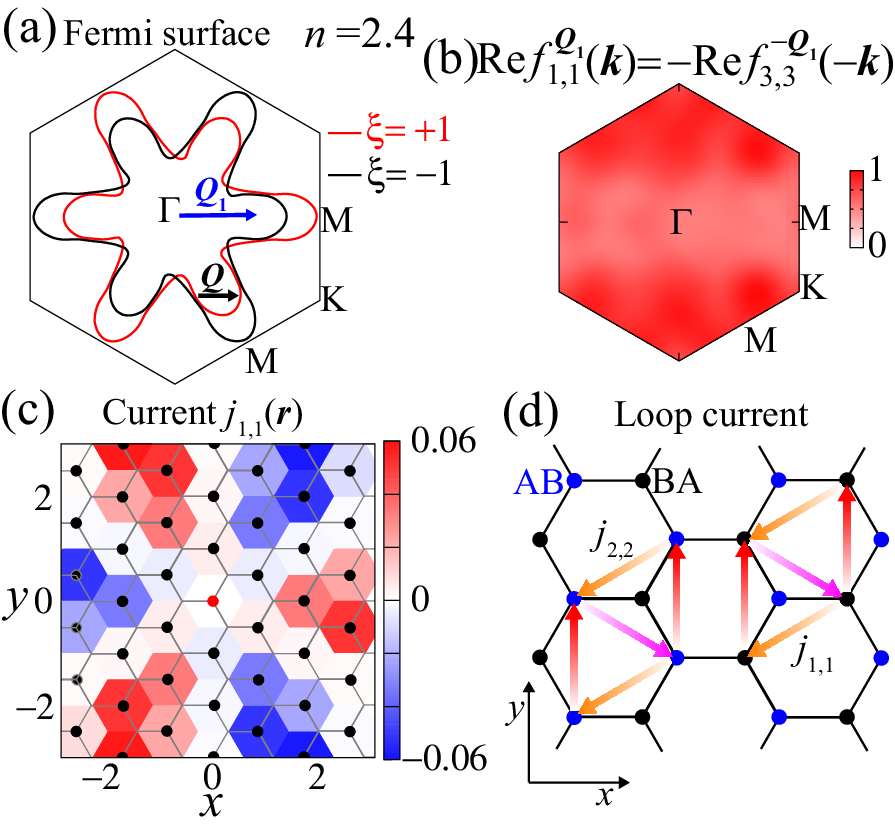}
\caption{
(a) FSs for $n=2.4$.
(b) Obtained form factor $f_{1,1}^{\bm{Q}_1}(\k)$, which breaks the time reversal
 symmetry. The intra- (inter-) valley relation Re$f_{1,1}^{\Q_1}(\k)$
 $={\rm Re}f_{2,2}^{\Q_1}(\k)$ (Re$f_{1,1}^{\Q_1}(\k)$ $=-{\rm
 Re}f_{3,3}^{-\Q_1}(-\k)$) is satisfied.
(c) $C_3$-symmetry current $j_{1,1}(\r)$ in the unit of $1$
meV/$\hbar$ without the
 form factor, where value at the BA spots (black dots) are depicted. Origin of $\r$
 is represented by red dot. $j_{2,2}(\r)$ for the AB spots is the same
 as $j_{1,1}(\r)$ for the BA spots.
(d) Schematic picture of the loop current $j_{1,1}$ and
 $j_{2,2}$ between the nearest intra-orbital sites.
In the valley polarized state, the charge loop current emerges.
}
\label{fig:n3.2}
\end{figure}

Finally, we discuss the time-reversal-symmetry-broken order in the present theory at
$n=2.4$.
Figures \ref{fig:n3.2}(a) and (b) show the FSs and obtained form factor
Re$f_{1,1}^{\Q_1}(\k)$, respectively. Here, the relation $\Q_1\sim 2\Q$ holds.
The obtained $\hat{f}^{\Q_1}(\k)$ violates time-reversal-symmetry
relation $f^{\Q_1}_{1,1}(\k)=f^{-\Q_1*}_{3,3}(-\k)$, and
brings the valley polarization.  This state is caused by the cooperation among the inter-valley Hartree term and the
 AL type quantum
interference with
$\Q=\Q'$ in Fig. \ref{fig:FS}(a). $\q$ dependence of $\lambda_{\q}$ is shown in SM F \cite{SM}.
The relation $U'=U$ $(J=0)$ is
important to realized the valley polarization.

%%%%%%%%%%%%%%%%%%%%%%%%%%%%%%%%%
%%%%%%%%%%%%%%%%%%%%%%%%%%%%%%%%%

We stress that interesting charge loop current
emerges in the valley polarized phase. In the absence of valley
polarization, the current from the orbital $m$ at $\r'$
 to the orbital $l$ at $\r$ is given as $j_{l,m}(\r,\r')=2{\rm Im}\langle
\tilde{h}^{0}_{l,m}(\r,\r')c^\dagger_l(\r)c_m(\r')
\rangle_{0}$, where $\tilde{h}^{0}_{l,m}(\r,\r')$ is the original hopping
integral \cite{Kontani-sLC}.
Figure \ref{fig:n3.2}(c) shows the current from the center site,
$j_{1,1}(\r)\equiv j_{1,1}(\r,\bm{0})$ due to the imaginary hopping
integrals in $\tilde{h}^0$. Figure \ref{fig:n3.2} (d) shows a schematic intra-orbital current pattern. 
The direction of rotation in
loop current $j_{1,1}$ is opposite to that in $j_{2,2}$.
Because of the relation $j_{1,1}(\r)=-j_{3,3}(\r)$, the charge loop current
is canceled, while pure valley loop current appears. 
In the presence of the valley polarization, the valley current is converted to the net charge loop
current. The magnetic
flux emerges in proportion to the valley polarization, and it will be measurable by several experimental methods.

In summary, 
we studied the origin of the nematic state in MATBG.
We found that the $\q=\0$ $C_3$-symmetry-breaking nematic
state near the VHS filling is identified as
the nematic bond order.
This order originates from  
prominent quantum interference among moderate \color{black}fluctuations
of $SU(4)$ valley+spin composite operators. \color{black}
This nematicity is robust once three VHS points
exist in each valley, insensitively to the shape of the FS.
% since the origin of nematic state is simplified as the intra-VHS
% attractive and inter-VHS repulsive interactions by using the simple
% three VHS model. 
We also found the emergence of the time-reversal-symmetry-broken valley
polarization, which accompanies the novel charge loop current. The present study revealed
unexpected interesting similarities and differences 
between MATBG and Fe-based superconductors.

\color{black}
In SM G \cite{SM}, 
we analyze the spin-channel DW equation \cite{Kontani-sLC},
and it is verified that the spin-channel instability is 
not magnified by the spin-channel VCs.
Thus, the charge-channel nematic bond order 
due to the charge-channel VCs is realized robustly.
In addition, in SM H \cite{SM},
we analyze the effect of the off-site Coulomb interactions $V_n$
included in the Kang-Vafek model \cite{Kang}.
It is verified that the nematic bond order 
driven by the valley + spin interference mechanism
(in $U$-only model) is stabilized 
in the presence of $V_n$.
\color{black}

%%%%%%%%%%%%%%%%%%%%%
\acknowledgements
We are grateful to 
%M. Koshino and
 Y. Yamakawa
for useful discussions.
This work was supported
by Grants-in-Aid for Scientific Research from MEXT,
Japan (No. JP19H05825, No. JP18H01175, and No. JP17K05543)

%%%%%%%%%%%%%%%%%%%%%%%
%\appendix
%\section{Supplemental Material}

%%%%%%%%%%%%%%%%%%%%%%%%
%references
%%%%%%%%%%%%%%%%%%%%%%%%

%\end{document}
%%%%%%%%%%%%%%%%%%%%%%%%%%%%%%%%%%%%%%%
\clearpage

\makeatletter
\renewcommand{\thefigure}{S\arabic{figure}}
\renewcommand{\theequation}{S\arabic{equation}}
\makeatother
\setcounter{figure}{0}
\setcounter{equation}{0}
\setcounter{page}{1}
\setcounter{section}{1}

\begin{widetext}
\begin{center}
{\bf 
[Supplementary Material] \\
$SU(4)$ Valley + Spin Fluctuation Interference Mechanism for Nematic
 Order in Magic Angle Twisted Bilayer Graphene: Impact of Vertex Corrections
}%
\end{center}

\begin{center}
Seiichiro Onari and Hiroshi Kontani
\end{center}

\begin{center}
\textit{Department of Physics, Nagoya University, Nagoya 464-8602, Japan}
\end{center}

\end{widetext}
\subsection{A: Model Hamiltonian of MATBG, formalism of the RPA and the DW equation}
\begin{figure}[!htb]
\includegraphics[width=.99\linewidth]{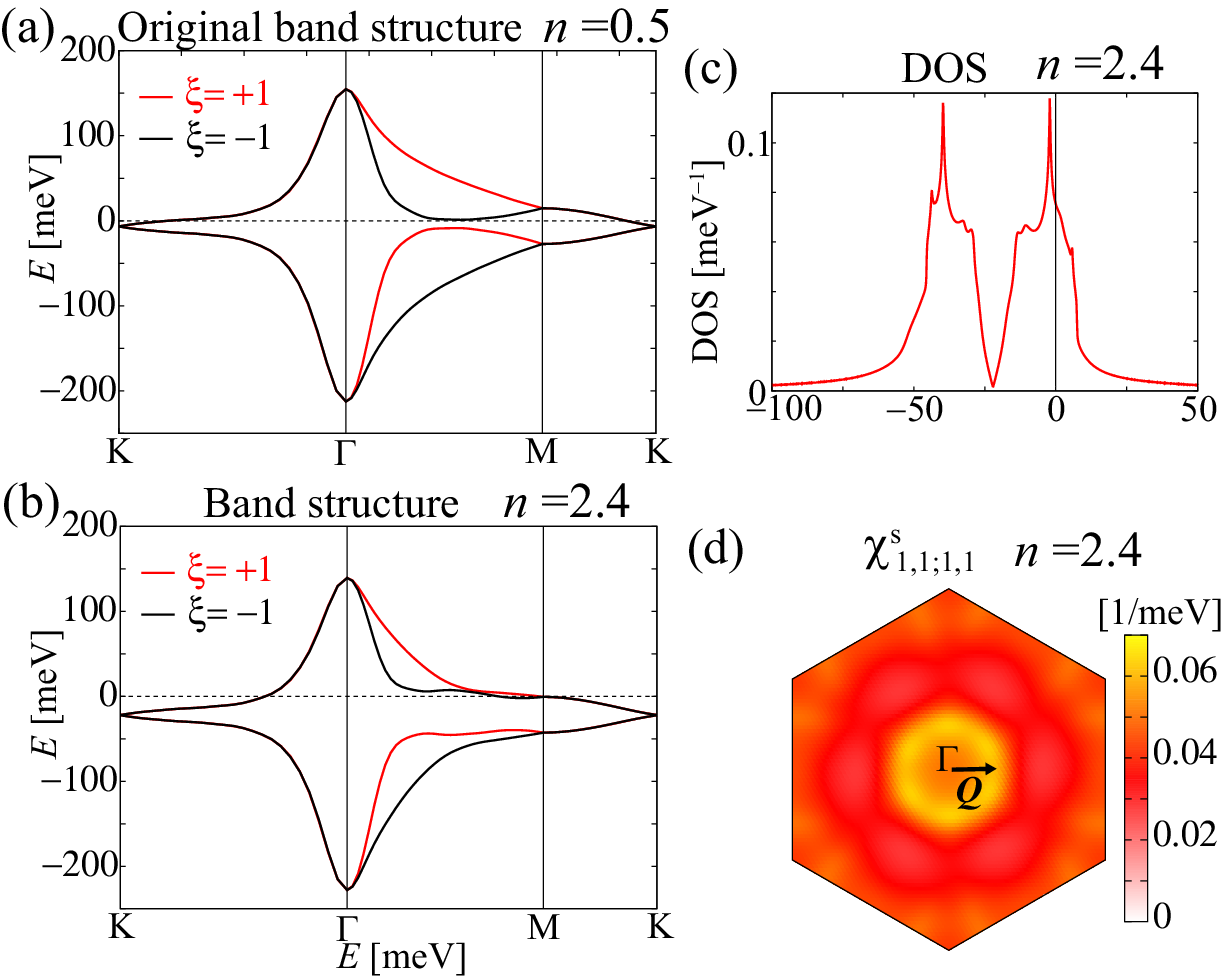}
\caption{
(a) Band
 structure in the original first-principles model for $n=0.5$, where red (black) lines denote valley $\xi=+1(-1)$.
(b) Band
 structure in the present model for $n=2.4$.
(c) DOS and (d) $\chi^c_{1,1;1,1}(\q,0)$ for $n=2.4$.
Arrow $\bm{Q}$ denotes the
 nesting vector.
}
\label{fig:FS-n3.2}
\end{figure}

First, we introduce model for MATBG by referring 
the first-principles tight-binding model in Ref. \cite{S-Koshino}.
However, in this original model, the VHS appears near $n\sim 0.7$, which is
different from the experimentally observed $n_{\rm VHS}\sim 2$
\cite{S-nematic-TBG}.
Figure \ref{fig:FS-n3.2}(a) shows the band structure of the original
model for $n=0.5$, where all the hopping integrals are magnified $50$ times in order to fit
the bandwidth observed by the STM
measurement \cite{S-nematic-TBG}.
To shift the VHS filling to the experimental one,
we reduce the magnitude of the imaginary part of second-nearest 
intra-orbital hopping 0.097 meV to 0.03 meV, while the real part is fixed.
Also, we reduce the magnitude of the imaginary part of
fourth-nearest 
intra-orbital hopping 0.039 meV to 0.02 meV.
Finally, we magnify all the hopping integrals $50$ times.
Figures \ref{fig:FS-n3.2}(b), (c), and (d) show the band
structure, the DOS, and $\chi^c_{1,1;1,1}(\q,0)$ in the obtained model,
respectively. At $n=2.4$, the Fermi energy $(E=0)$ is above
the energy of VHS.
Although the band structure in the present model is similar to that in the original
model, the energy difference between the valleys near the Fermi energy
along $\Gamma$-M line increases in the present model.

%%%%%%%%%%%%%%%%%%%%%%%%%%%%%%%%%
%%%%%%%%%%%%%%%%%%%%%%%%%%%%%%%%%

Here, we explain the Coulomb interaction in the orbital basis introduced in the present study.
\color{black}Only the Coulomb interactions between the orbitals with the same center
position are
taken into account.
The Coulomb interactions for the spin and charge channels
in the main text are generally given as
\begin{equation}
(\Gamma^{\mathrm{s}})_{l_{1},l_{2};l_{3},l_{4}} = \begin{cases}
U, & l_1=l_2=l_3=l_4 \\
U', & l_1=l_3= l_2\pm2=l_4\pm2 \\
J, & l_1=l_2= l_3\pm2=l_4\pm2 \\
J, & l_1=l_4= l_2\pm2=l_3\pm2 \\
0 , & \mathrm{otherwise},
\end{cases}
\end{equation}
\begin{equation}
({\hat \Gamma}^{\mathrm{c}})_{l_{1},l_{2};l_{3},l_{4}} = \begin{cases}
-U, & l_1=l_2=l_3=l_4 \\
U' - 2J , & l_1=l_3= l_2\pm2=l_4\pm2 \\
-2U' + J , & l_1=l_2= l_3\pm2=l_4\pm2 \\
-J, & l_1=l_4= l_2\pm2=l_3\pm2 \\
0 . & \mathrm{otherwise}.
\end{cases}
\end{equation}

Hamiltonian of the Coulomb interaction is given as
\begin{eqnarray}
H'=-\!\!\!\!\sum_{\k\bm{k'}\q,\sigma\sigma'}\sum_{l_1l_2l_3l_4}&&\left(\frac{\Gamma^c+\Gamma^s\sigma\sigma'}{4}\right)_{l_1,l_2;l_3,l_4}\nonumber\\
&&\times c^{l_1\dagger}_{\k+\q,\sigma}c^{l_2}_{\k,\sigma}c^{l_3\dagger}_{\bm{k'}-\q,\sigma'}c^{l_4}_{\bm{k'},\sigma'},
\end{eqnarray}
where $\sigma,\sigma'=\pm 1$ denote spin.

We set $U'=U$ and $J=0$ in the present study, since these relations are
verified when only the Coulomb interactions between the orbitals with the
same center position are taken into account\cite{S-Koshino,S-Markus}. 
First, the relation $U'=U$ is verified by the fact that the
density of orbital 1(2) is the same as that of orbital 3(4) since the wave
function of orbital 1(2) is identical to the complex conjugate wave
function of orbital 3(4). Next, the inter-valley exchange interaction
$J/U\ll 1$ has been explained in Ref. \cite{S-Koshino}, where the
integration in the calculation of the exchange interaction becomes very
small due to the strongly oscillating phase. Moreover, the relation
$J/U<10^{-3}$ has been confirmed in Ref. \cite{S-Markus}. \color{black}
The effect of the inter-site Coulomb interaction is discussed in SM H. \color{black}
%Here, we drop the inter-site Coulomb interaction
%$V_{\r,\r'}$, because of the following reasons;
%In the RPA, the spin fluctuations and valley fluctuations are not
%affected by $V_{\r,\r'}$.
%Moreover, $\lambda_{\q}$ with the nearest neighbor
%$V_{\r,\r'}<\frac{U}{2}$ $(\alpha_c<0.95)$ is almost the same as that without $V_{\r,\r'}$.
%Therefore, we neglect $V_{\r,\r'}$ in the main text, by expecting the
%screening effect on $V_{\r,\r'}$.
%\color{black}
%Thus, the Coulomb interaction in the main text corresponds to the
% Kang-Vafek model\cite{S-Kang} without inter honeycomb site interaction,
% which satisfies the
% hoppinglike terms $\alpha_1=0$ in Eq.(9) and
% $V^{(0)}_{j,p;j',p'}\propto\delta_{jj'}\delta_{p,p'}$ in Eq.(S6) in Ref. \cite{S-Kang}.
%\color{black}

By using the multiorbital Coulomb interaction,
the spin (charge) susceptibility in the RPA is given by  
\begin{equation}
{\hat \chi}^{s(c)}(q)={\hat\chi^0}(q)[1-{\hat \Gamma}^{s(c)}{\hat
\chi^0(q)}]^{-1},
\end{equation}
where the irreducible susceptibility is
\begin{equation}
\chi^0_{l,l';m,m'}(q)= -\frac{T}{N}\sum_k
G_{l,m}(k+q)G_{m',l'}(k).
\end{equation}
${\hat G}(k)$ is the multiorbital Green function without self-energy 
${\hat G}(k)=[(i\e_n+\mu){\hat1}-{\hat{h}}^0(\k)]^{-1}$ 
for $k=[\k,\e_n=(2n+1)\pi T]$. 
Here, ${\hat{h}}^0(\k)$ is the matrix expression of $H^0$ 
and $\mu$ is the chemical potential.

The spin (charge) Stoner factor $\alpha_{s(c)}$ 
is defined as
the maximum eigenvalue of $\hat{\Gamma}^{s(c)}\hat{\chi}^0(\bm{q},0)$.
$\a_c=\a_s=\a$ is satisfied due to the relation $U'=U$ $(J=0)$.
In the present study, we fix $\a=0.83$ and $T=1.5$ meV, by setting $U=39$
$(42)$ meV for $n=2.0$ $(2.4)$. ($U$ is the solo parameter in the present study.)
We use $N=72\times 72$ $\k$ meshes and $512$ Matsubara frequencies.

The charge-channel irreducible interaction $\hat{I}^{c,\bm{q}}(k,k')$
in the DW equation (\ref{eqn:linearized}) \cite{S-Onari-form,S-Onari-B2g} is given by
\begin{eqnarray}
&& 
I^{c,\bm{q}}_{l,l';m,m'}(k,k')=\sum_{b=s,c}
\left[-\frac{a^b}{2} V^{b}_{l,m;l',m'}(k-k')\right.
\nonumber \\
&& 
+\frac{T}{N}\!\!\sum_{p,l_1,l_2,m_1,m_2}\!\!\!\!\!\!\!\!\!\!
 \frac{a^b}{2} V^{b}_{l,l_1;m,m_2}\left(p+\frac{\q}{2}\right)V^{b}_{m_1,l';l_2,m'}\left(-p+\frac{\q}{2}\right)
 \nonumber \\
&& \qquad\qquad
\times G_{l_1,m_1}(k-p)G_{l_2,m_2}(k'-p)
\nonumber \\
&&
+\frac{T}{N}\!\!\!\!\sum_{p,l_1,l_2,m_1,m_2}\!\!\!\!\!\!\!\!\!\!
 \frac{a^b}{2} V^{b}_{l,l_1;l_2,m'}\left(p+\frac{\q}{2}\right)V^{b}_{m_1,l';m,m_2}\left(-p+\frac{\q}{2}\right)
 \nonumber \\
&& \qquad\qquad
\left.\times G_{l_1,m_1}(k-p)G_{l_2,m_2}(k'+p)\right],
%-({\rm Double\;counting\;} [\hat{\Gamma}^{s(c)}]^2 \;{\rm terms})
\label{eqn:S-K} 
\end{eqnarray}
%\begin{eqnarray}
%&& \!\!\!\!\!\!\!\!\!\!\!
%K^{\bm{q}}_{l,l';m,m'}(k,k')=
%\sum_{l'',m''}\left[\frac32 V^{s}(k-k')+\frac12 V^{c}(k-k')\right]_{l,l'';l',m''}
%\nonumber \\
%&& \ \ \ \ \ \ \ \ \ \ 
%\times G^0_{m',m''}\left(k'-\frac{\q}{2}\right)G^0_{l'',m}\left(k'+\frac{\q}{2}\right)
% \nonumber \\
%&&
%-\frac{T}{N}\!\!\!\!\sum_{p,l_1,l_2,l_3,m_1,m_2,m_3}\!\!\!\!\!\!\!\!\!\!\!\!\!\!
% \left[ \frac32 V^{s}_{l_1,l_2;l',m_1}\left(p-\frac{\q}{2}\right)V^{s}_{l,l_3;m_2,m_3}\left(p+\frac{\q}{2}\right) \right.
% \nonumber \\
%&& 
%\left. +\frac12 V^{c}_{l_1,l_2;l',m_1}\left(p-\frac{\q}{2}\right)V^{c}_{l,l_3;m_2,m_3}\left(p+\frac{\q}{2}\right) \right]G^0_{l_3,m_1}(k-p)
%\nonumber \\
%&& 
%\times \left[{\Lambda}^\q_{m',l_1;l_2,m_3;m_2,m}(k';p)+{\Lambda}^\q_{m',m_3;m_2,l_1;l_2,m}(k';-p)\right],
%%-({\rm Double\;counting\;} [\hat{\Gamma}^{s(c)}]^2 \;{\rm terms})
%\label{eqn:S-K} 
%\end{eqnarray}
%
where $a^s=3$, $a^c=1$, $p=(\p,\w_l)$, and
$\hat{V}^{s(c)}(q)=\hat{\Gamma}^{s(c)}+\hat{\Gamma}^{s(c)}\hat{\chi}^{s(c)}(q)\hat{\Gamma}^{s(c)}$.
%in Ref. \cite{Onari-SCVC,Onari-SCVCS}.
\color{black}In the conserving
approximation scheme\cite{S-Luttinger,S-Baym}, $\hat{I}^{c,\q=\bm{0}}(k,k')\equiv{\hat I}^{\uparrow\uparrow,\q=\bm{0}}+{\hat I}^{\uparrow\downarrow,\q=\bm{0}}$
is given by 
\begin{equation}
I^{\sigma\sigma',\q=\bm{0}}_{l,l';m,m'}(k,k')=\frac{\delta^2\Phi}{\delta
 G^\sigma_{l',l}(k)\delta G^{\sigma'}_{m,m'}(k')},
\end{equation}
where $\Phi$ is the Luttinger-Ward function within the one-loop
approximation, and $\sigma$ denotes spin. \color{black}
In Eq. (\ref{eqn:S-K}),
the first line corresponds to the Maki-Thompson (MT) term,
and the second and third lines give the AL1 and AL2 terms, respectively.
\color{black}The AL terms are enhanced by the fluctuation interference
$\hat{\chi}^{s(c)}(\Q)\times\hat{\chi}^{s(c)}(\bm{Q'})$ shown in Fig
\ref{fig:FS}(a). Thus, $\q=\Q+\bm{Q'}=\bm{0}$ nematic bond order is due to
fluctuation interference with
$\bm{Q'}=-\Q$. \color{black}
In the MT term,
the first-order term with respect to ${\hat{\Gamma}}^{s,c}$ 
gives the Hartree--Fock (HF) term in the mean-field theory.

%%%%%%%%%%%%%%%%%%%%%%%%%%%%%%%%%%%%%%
\color{black}
\subsection{B: Fifteen-channel valley+spin fluctuations in the MATBG model}

In the main text, we explained that the 
nematic bond order in MATBG originates from the 
valley+spin fluctuation interference mechanism.
Here, we explain why the present mechanism 
gives rise to remarkable nematic instability in MATBG, even stronger than Fe-based and cuprate superconductors.

We introduce the Pauli matrices for the spin-channel ($\rho=\pm1$)
and the valley-channel ($\xi=\pm1$), 
${\hat \sigma}_m$ and ${\hat \tau}_n$, respectively.
Here, $m,n=1,2,3$.
Then, the on-site Coulomb interaction is expressed as \cite{S-Tazai-CeB6}
\begin{eqnarray}
H'&=&\frac{U}{16} \sum_{i}\left[ -\sum_{\mu,\nu}({\hat O}^i_{\mu,\nu})^2 +4({\hat O}^i_{0,0})^2\right],
\label{eqn:HU}
\\
{\hat O}^i_{\mu,\nu}&=& \sum_{\rho,\xi} ({\hat \s}_\mu {\hat \tau}_\nu)_{\rho\xi,\rho'\xi'}c_{i\rho\xi}^\dagger c_{i\rho'\xi'},
\label{eqn:O-def}
\end{eqnarray}
where $\mu,\nu=0\sim3$, $i$ is site index, and ${\hat \s}_0$ (${\hat \tau}_0$)
is the identity matrix for spin (valley) sector. 
The Coulomb interaction $H'$ in Eq. (\ref{eqn:HU}) apparently
possesses $SU(4)$ symmetry.

Next, we study the susceptibility with respect to the 
spin-valley operator ${\hat O}_{\mu,\nu}$ in Eq. (\ref{eqn:O-def}):
\begin{eqnarray}
\chi_{\mu,\nu}(\q)= \frac12 \int_0^\beta du
\langle T_u {\hat O}_{\mu,\nu}(\q,u) {\hat O}_{\mu,\nu}(-\q,0) \rangle,
\\
{\hat O}_{\mu,\nu}(\q)= \sum_{\k,a} ({\hat \s}_\mu {\hat \tau}_\nu)_{\rho\xi,\rho'\xi'}
c_{\k,a\rho\xi}^\dagger c_{\k+\q,a'\rho'\xi'}.
\end{eqnarray}
Here, $a=$AB or BA in Fig. \ref{fig:FS} (b),
and $u$ is the imaginary time.
Here, $\chi_{m,0}$ and $\chi_{0,n}$ 
are the spin and valley susceptibilities, respectively.
Also, $\chi_{m,n}$ represent the susceptibility of the 
``spin-valley quadrupole order'',
composed of the products of spin and pseudospin (= valley) operators.

%Except for $\mu=\nu=0$,
%each component of $\chi_{\mu,\nu}(\q)$ 
%is almost equivalently enlarged by $U$, 
%owing to the $SU(4)$ symmetric 
%Coulomb interaction in Eq. (\ref{eqn:HU}).

\begin{figure}[!htb]
\includegraphics[width=.8\linewidth]{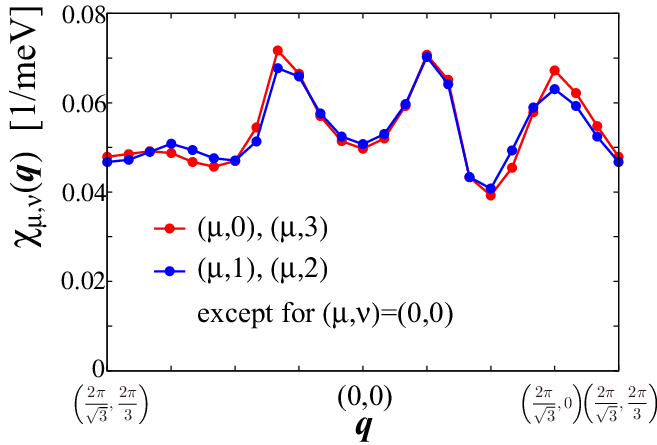}
\caption{
$\chi_{\mu,\nu}(\q)$ in the present MATBG model given by the RPA.
}
\label{fig:chiRPA16}
\end{figure}

Figure \ref{fig:chiRPA16} represents the fifteen susceptibilities
$\chi_{\mu,\nu}(\q)$ ($(\mu,\nu)\ne(0,0)$)
given by the RPA in the MATBG model.
By reflecting the $SU(4)$ symmetric 
Coulomb interaction in Eq. (\ref{eqn:HU}),
%approximate $SU(4)$ symmetry of the MATBG Hamiltonian,
all $\chi_{\mu,\nu}(\q)$ take very similar values:
Seven components with $(\mu,\nu)=(m,0),(\mu,3)$ are equivalent,
and eight components with $(\mu,\nu)=(\mu,1),(\mu,2)$ are 
also equivalent. Thus, following relations are exact,
\begin{eqnarray}
\chi_{m,0}(\q)&=&\chi_{\mu,3}(\q)=\chi^{(1)}(\q)\quad {\rm for\,\, any}\,m\,{\rm and }\,\mu, \nonumber \\
\chi_{\mu,1}(\q)&=&\chi_{\nu,2}(\q)=\chi^{(2)}(\q) \quad {\rm for\,\, any}\,\mu\,{\rm and }\,\nu,
\end{eqnarray}
where $\chi^{(1)}(\q)\sim\chi^{(2)}(\q)$ in the present MATBG model are shown in Fig. \ref{fig:chiRPA16}.
These approximate fifteen-fold eigenstates with the form factors
$f_{\mu\nu}\approx {\hat \s}_\mu {\hat \tau}_\nu$ are also obtained by
the DW equation analysis.
However, they do not correspond to the largest eigenvalue.
The largest eigenvalue in the DW equation is the 
charge-channel bond-order state with $E$-symmetry,
${\bm f}_{\rm bond}\approx {\hat \s}_0 {\hat \tau}_0 
(f^{\bm{0}}(\k),f'^{\bm{0}}(\k))$, as we derived in the main text.
The eigenvalue of this bond-order state is strongly magnified
by the AL-type VCs in the DW equation.

%%%%%%%%%%%%%%%%%%%%%%%%%%%%%%%%
%%%%%%%%%%%%%%%%%%%%%%%%%%%%%%%%%%%%%%

Now, we consider the main driving force of the nematic order.
The AL terms in Fig. \ref{fig:chi} (b) are composed of the convolutions of 
$\chi_{\mu,\nu}(\q)$.
Because fifteen susceptibilities in
Fig. \ref{fig:chiRPA16} exhibit similar $\q$-dependences,
the ratio of the contributions from the spin ($\chi_{m,0}(\q)$),
the valley ($\chi_{0,n}(\q)$), 
and the quadrupole ($\chi_{m,n}(\q)$) susceptibilities
to the AL terms are approximately 
$\frac{3}{15}$ : $\frac{3}{15}$ : $\frac{9}{15}$.
Therefore, the AL interference process is caused by
not only independent spin and valley fluctuations, 
but also spin+valley composite (quadrupole) fluctuations. 
%We define fifteen susceptibilities $\chi_l(\q)$ $(l=1\sim15)$ as $\chi_l(\q)=\chi_{l,0}(\q)$, $\chi_{m+3}(\q)=\chi_{0,m}(\q)$,
%and $\chi_{l*3+m+3}(\q)=\chi_{l,m}(\q)$ for $l=1\sim3$ and $m=1\sim3$.
\color{black}

%%%%%%%%%%%%%%%%%%%%%%%%%%%%%%%%%
\begin{figure}[h]
\includegraphics[width=.99\linewidth]{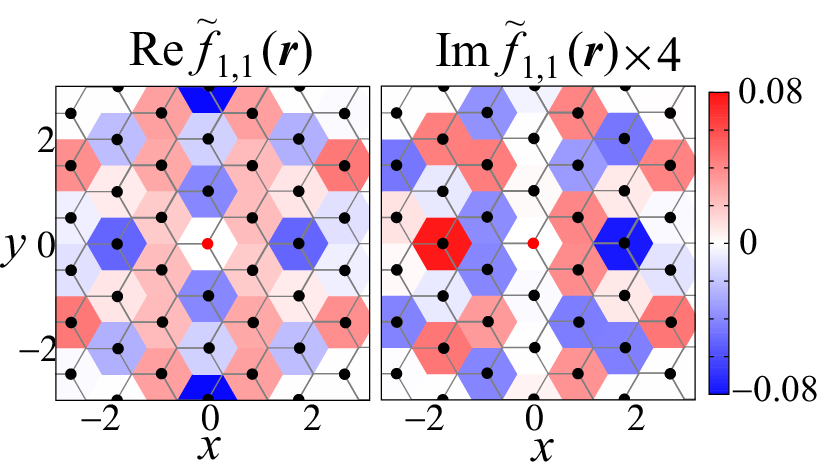}
\caption{
Re$\tilde{f}_{1,1}(\bm{r})$ and Im$\tilde{f}_{1,1}(\bm{r})$ of the
  form factor.
\color{black}Centers of
 orbital 1 (BA sites) are represented by black dots, and origins of $\bm{r}$ are shown by
 the red dots. Color maps show value at the black dots.\color{black}
}
\label{fig:fourier}
\end{figure}
%%%%%%%%%%%%%%%%%%%%%%%%%%%%%%%%%

\color{black}
\subsection{C: Parity mixing in form factor}
\begin{figure}[h]
\includegraphics[width=.99\linewidth]{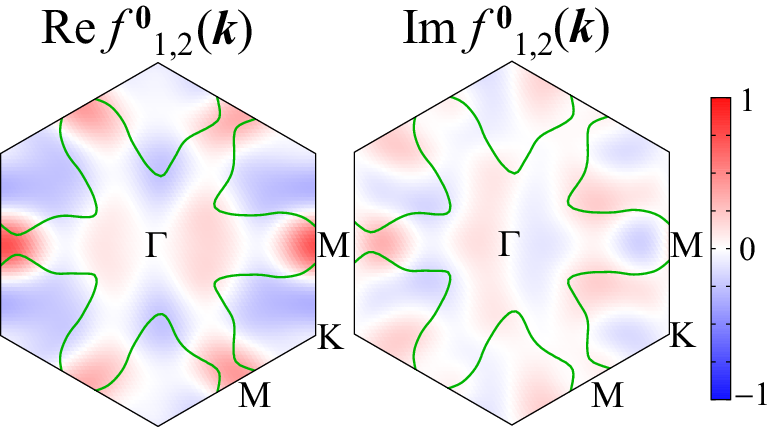}
\caption{
 Off-diagonal form factor explained in the main text for $n=2.0$
The green lines indicate FSs for $\xi=+1$. 
}
\label{fig:offdia}
\end{figure}

Here, we explain the real-space structure of the obtained $\q=\0$ form factor in
MATBG. Figure \ref{fig:offdia} shows real part of Fourier transformed
form factor Re$\tilde{f}_{1,1}(\r)$
and imaginary part of that Im$\tilde{f}_{1,1}(\r)$.
Re$\tilde{f}_{1,1}(\r)$ is even parity, and it gives the bond order between the position $\0$ and $\r$. On the other
hand, Im$\tilde{f}_{1,1}(\r)$  has odd parity. Thus, parity mixing in
the form factor is confirmed in MATBG, and it gives the
time-reversal-invariant valley current order \cite{S-Kontani-sLC}. Thus, both even-parity and odd-parity components coexist in the present order parameter.  This ``parity mixing order'' is a natural consequence of the violation of the inversion symmetry in MATBG: The primary even-parity component in $f$ induces sizable secondary odd-parity one through the imaginary intra-orbital hopping integrals due to the inversion symmetry breaking.

Figure \ref{fig:offdia} shows the off-diagonal
form factor $f^\0_{1,2}(\k)$, which is slightly smaller than
the diagonal form factor. This form factor is invariant under the time
reversal operation
$f^\0_{1,2}(\k)=f^{\0*}_{3,4}(-\k)$. Even parity Re$\tilde{f}_{1,2}(\r)(=$Re$\tilde{f}_{2,1}(-\r))$
is also mixed by odd parity Im$\tilde{f}_{1,2}(\r)(=-$Im$\tilde{f}_{2,1}(-\r))$.

\begin{figure}[h]
\includegraphics[width=.6\linewidth]{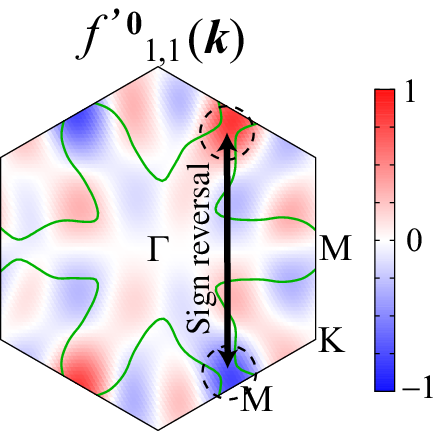}
\caption{
Degenerate form factor $f'^\0(\k)$ for $n=2.0$.
The green lines and black arrow indicates FSs for $\xi=+1$ and
 sign-reversal between the VHS points. 
}
\label{fig:deg}
\end{figure}

Since the obtained form factors belong to the two-dimensional $E$
representation, the other form factors
$\hat{f}'^{\0}$ shown in Fig. \ref{fig:deg} has
the same eigenvalue $\lambda_\0$. 
The direction of anisotropy of $f'^{\bm{0}}$ is different from that of
$f^{\bm{0}}$ shown in the main
text. Thus, we can rotate the direction of anisotropy by making the
linear combination of $\hat{f}^{\0}$ and $\hat{f}'^{\0}$.
In real systems, the direction of nematicity will be fixed by the anharmonic phonons and/or the fourth-order terms in the Ginzburg-Landau free energy.
\color{black}
%%%%%%%%%%%%%%%%%%%%%%%%%%%%%%%%%
%%%%%%%%%%%%%%%%%%%%%%%%%%%%%%%%%

%%%%%%%%%%%%%%%%%%%%%%%%%%%%%%%%%
%%%%%%%%%%%%%%%%%%%%%%%%%%%%%%%%%
\color{black}
\subsection{D: Robustness of nematic order and impact of three VHS
  points}
\subsubsection{1. Analytic discussion on a simple three VHS model}
First, we explain an origin of strong intra- and inter-VHS interactions
due to the VCs in the DW Eq. (\ref{eqn:linearized}) in MATBG.
 The derived interaction originates from the fluctuation interference mechanism, and it naturally promotes the nematicity in $E$-symmetry.

%%%%%%%%%%%%%%%%%%%%%%%%%%%%%%%%%
\begin{figure}[h]
\includegraphics[width=.9\linewidth]{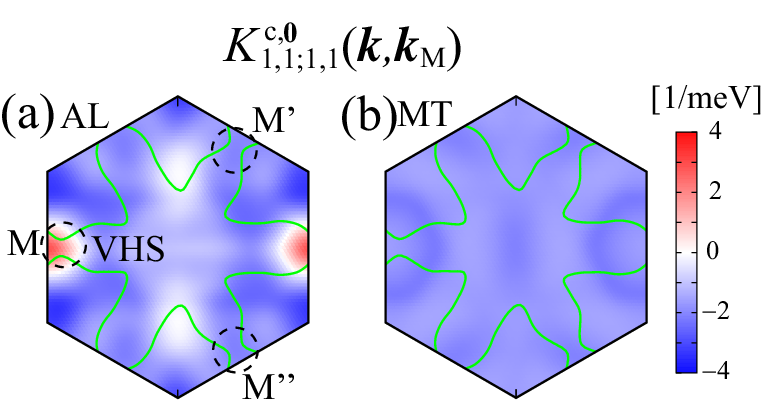}
 \caption{\color{black}
(a) $\bm{k}$ dependences of kernel
 $K^{c,\bm{0}}_{1,1;1,1}(\k,\k_M)$ for the AL terms and (b) that for the MT term, composed
 of orbital $1$. Here,
 $\k_M=(-\frac{2\pi}{\sqrt{3}},0)$, $\k_{M'}=(\frac{\pi}{\sqrt{3}},\pi)$, and
 $\k_{M''}=(\frac{\pi}{\sqrt{3}},-\pi)$.
}
\label{fig:Kernel}
\end{figure}
%%%%%%%%%%%%%%%%%%%%%%%%%%%%%%%%%

Figures \ref{fig:Kernel}(a) and (b) show $\bm{k}$ dependences of kernel
 $K^{c,\bm{0}}_{1,1;1,1}(\k,\bm{k}_M)$ for the AL and MT terms composed of the
 orbital 1 in valley $\xi=+1$, where $\k_M=(-\frac{2\pi}{\sqrt{3}},0)$ at the M
 point. The obtained kernel takes sizable positive value at $\k\sim \k_M$,
 while it exhibits negative values at $\k\sim\k_{M'}$ and $\k_{M''}$.
 The $\k$ dependence of
 $K^{c,\bm{0}}_{1,1;1,1}(\k,\bm{k}_M)$ favors the $E$-symmetry nematicity.
In the following, we explain that $\bm{k}$ dependence of
 $K^{c,\bm{0}}_{1,1;1,1}(\k,\bm{k}_M)$ is mainly
 caused by the AL1 term. We note that $K^{c,\bm{0}}(\k,\bm{k'})$
 is proportional to $I^{c,\bm{0}}(\k,\bm{k'})$.

\begin{figure}[h]
\includegraphics[width=.99\linewidth]{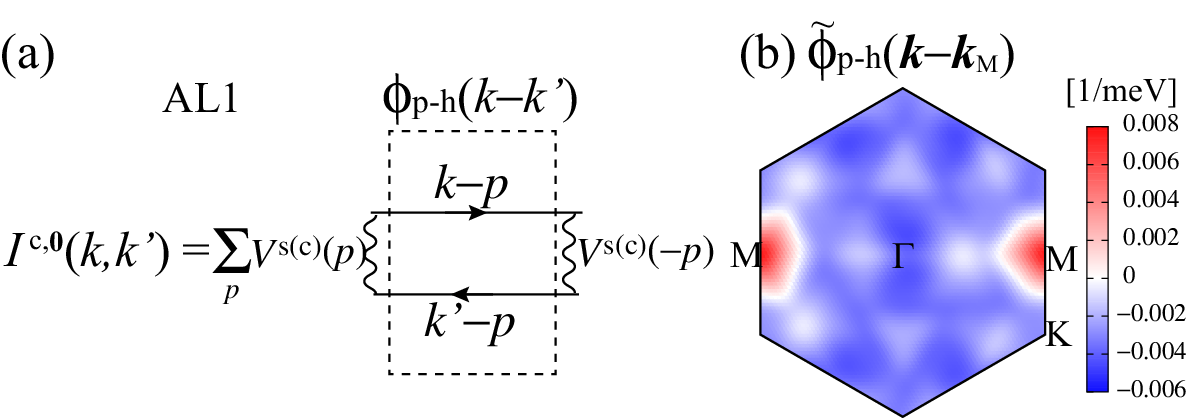}
 \caption{\color{black}
(a)Feynman diagram of $\phi_{\mbox{p-h}}(k-k')$ in $I^{c,\bm{0}}(k,k')$ of
 the AL1 term.
(b)$\bm{k}$ dependence of $\tilde{\phi}_{\mbox{p-h}}(\k-\bm{k}_M)$ for
 cutoff $\pi T<\w_c<3\pi T$. Positive peak appears at $\k=\bm{k}_M$.
}
\label{fig:phi}
\end{figure}

As we discussed in Ref. \cite{S-Onari-B2g}, the $k$, $k'$
dependence of AL1 term is mainly determined by the momentum dependence
of the particle-hole (p-h) propagator, rather than $q$ dependence of $\chi^{s(c)}(q)$.
The p-h propagator is given as
\begin{eqnarray}
&&\phi_{\mbox{p-h}}(\q,\w_m)\equiv \nonumber\\
&&\frac{T}{N}\displaystyle\sum_{\p,\w_l}
 G_{1,1}(\q+\p,\w_m+\w_l)G_{1,1}(\p,\w_l)\nonumber\\
&&\times\theta(\w_c-|\w_m+\w_l|)\theta(\w_c-|\w_l|),
\label{Eq:phi}
\end{eqnarray}
and the AL1 term in $I^{c,\bm{0}}(k,k')$ in
 Eq. (\ref{eqn:S-K}) is proportional to $\phi_{\mbox{p-h}}(k-k')$ since $\q$ dependence of $\chi^{s(c)}(q)$ is
 moderate in MATBG. 
The cutoff energy
 $\w_c \ll E_{\rm F}$ is related to energy-scale of spin (valley) fluctuations ${\hat \chi}^{s(c)}$ in
 $\hat{V}^{s(c)}$ in AL1 term. 
Feynman diagram of $\phi_{\mbox{p-h}}(q)$ is shown in
 Fig. \ref{fig:phi}(a).

Since the energy scale $\w_s$ of ${\hat \chi}^{s(c)}(q)$ satisfies
$\w_s\lesssim 2\pi T$ in the moderately correlated MATBG, the relation $\pi
T<\w_c<3\pi T$ in Eq. (\ref{Eq:phi}) is reasonable.
For simplicity, we apply the lowest Matsubara approximation for the
kernel function as $I^{c,\bm{0}}(\k,\pi T, \bm{k'},\pi T)+ I^{c,\bm{0}}(\k,\pi T,\bm{k'},-\pi T)$, which is proportional to 
\begin{eqnarray}
\tilde{\phi}_{\mbox{p-h}}(\q)\equiv{\phi}_{\mbox{p-h}}(\q,0)+{\phi}_{\mbox{p-h}}(\q,
 2\pi T).
\end{eqnarray}
Figure \ref{fig:phi}(b) shows the obtained
 $\tilde{\phi}_{\mbox{p-h}}(\k-\k_M)$ as a function of $\k$. We see that
 $\tilde{\phi}_{\mbox{p-h}}(\k-\k_M)$ well reproduces 
 the result for the AL terms in Fig. \ref{fig:Kernel}(a).

Hereafter, we explain an origin of the $\q$ dependence of
 $\tilde{\phi}_{\mbox{p-h}}(\q)$.
It is approximately given as
\begin{eqnarray}
\!\!\!\!\tilde{\phi}_{\mbox{p-h}}(\q)&\sim& \frac{T}{N}\sum_{\bm{p}}\frac{2\varepsilon_{\q+\bm{p}}\varepsilon_{\p}}{[(\pi
 T)^2+\varepsilon_{\q+\bm{p}}^2][(\pi
 T)^2+\varepsilon_{\bm{p}}^2]},
\label{Eq:analy}
\end{eqnarray}
where $\varepsilon_{\k}=h^0_{1,1}(\k)-\mu$.
Thus, $\tilde{\phi}_{\mbox{p-h}}(\q)$ exhibits sizable positive value for $\q=
\bm{0}$, while it becomes small in magnitude and can be negative for finite $|\q|$ due to the cancellation of
$\varepsilon_{\q+\bm{p}}\varepsilon_{\p}$ in the numerator of Eq. (\ref{Eq:analy}).
 
The AL1 term $I^{c,\bm{0}}(k,k')$ is proportional to
 $\tilde{\phi}_{\mbox{p-h}}(\k-\bm{k'})$, and it is enhanced by the
 convolution of valley+spin
 susceptibilities at low temperatures \cite{S-Onari-B2g}.
In addition, as shown in Fig. \ref{fig:Kernel}(b), the MT term $(\propto
 -V^s_{1,1;1,1}(k-k'))$ gives negative interaction. 

We stress that $\tilde{\phi}_{\mbox{p-h}}(\q)$ becomes always negative
if we set $\w_c\sim E_F$. In fact,
$\tilde{\phi}_{\mbox{p-h}}(q)=-\chi^0(\q,0)-\chi^0(\q,2\pi T)$ for
$\w_c=\infty$. Therefore, the relation $\w_c\ll E_F$ is important to
obtain the nematic order.

%%%%%%%%%%%%%%%%%%%%%%%%%%%%%%%%%
%%%%%%%%%%%%%%%%%%%%%%%%%%%%%%%%%

Next, we clarify that the $E$-symmetry nematicity is caused by the intra-VHS interaction $I>0$
 and inter-VHS one $I'<0$ on the simple three VHS
 model in the main text. In this model, $\k$ and $\bm{k'}$ point are
 limited to the three VHS points $A$, $B$, and $C$ in
 Fig. \ref{fig:Sigma}(c). Thus, $I^{c,\bm{0}}_{1,1;1,1}(\k,\bm{k'})$ in the DW Eq. is given by the following
 $3\times 3$ matrix,
\begin{equation}
\hat{I}^{c,\bm{0}}_{1,1;1,1}=
\left(
\begin{array}{ccc}
I& I'& I'\\
I'& I& I'\\
I'& I'& I
\end{array}
\right).
\end{equation}
By using the DOS of orbital 1 at the Fermi energy $N_0$ and the form factors
$\vec{f}=^t\!\!(f_A,f_B,f_C)$, where 
$f_X$ denotes the form factor
 for orbital 1 at the VHS $X(=A,B,C)$ point, the DW equation in
 Eq. (\ref{eqn:linearized}) is given as
\begin{equation}
\lambda_{\bm{0}}\vec{f}=N_0\hat{I}^{c,\bm{0}}_{1,1;1,1}\vec{f}.
\end{equation}

The solved eigenvalues are doubly degenerate $\lambda_E \propto I-I'$ and non-degenerate
$\lambda_A \propto I+2I'$. The former $\lambda_E$
has degenerate form factor (eigenvector)
$(f_A,f_B,f_C)\propto(1,1,-2)/\sqrt{6}$, $(1,-1,0)/\sqrt{2}$.
These form factors correspond to $E$-symmetry $f^{\bm{0}}_{1,1}$ in Fig. \ref{fig:Sigma}(b) and
$f'^{\bm{0}}_{1,1}$ in SM C.
The latter $\lambda_A$ has $(f_A,f_B,f_C)\propto(1,1,1)/\sqrt{3}$,
which corresponds to the  $A$-symmetry order. 
Thus, the nematic $E$-symmetry with $\lambda_E(>\lambda_A)$ is explained by
the three VHS model due to the relations $I>0$ and $I'<0$.  The relation
$I'<0$ induces sign-reversal in form factors between the VHS points.
The nematic order due to the valley+spin fluctuation interference is
robust independently of the shape of FS and position of VHS point once
three VHS points exist in each valley. 

 To summarize, we studied the charge-channel interaction in the DW Eq. (\ref{eqn:linearized}), 
$I^{c,\bm{0}}$, beyond the mean-field approximation.
Strong attractive intra-VHS interaction emerges due to the valley+spin fluctuation interference described by the AL processes.
In addition, repulsive inter-VHS interaction originates from the AL + MT processes.
These intra- and inter-VHS interactions naturally induce the $E$-symmetry nematicity in MATBG.
The nematic order is robust once three VHS points exist in each valley.

%\subsection{C: Nematic state obtained by the original Koshino model}

%\subsection{D: Robustness of the nematic bond ordered state}

\color{black}
\subsubsection{2. Numerical study on the original first-principles MATBG model}
%
%In the present study, we introduced minimum additional terms into the original Koshino
%model \cite{S-Koshino} in order to realize the experimental VHS
%filling $n_{\rm VHS}\sim 2$ \cite{S-nematic-TBG}.
%Based on the modified model explained in SM A, we explain  the emergence of
% the nematic bond order in the MATBG in the main text. The obtained nematic bond order is
%consistent with experimental results \cite{S-nematic-TBG}.
%We clarify that the origin of the nematicity is the valley+spin
%fluctuation interference due to the AL-type VC.

\begin{figure}[h]
\includegraphics[width=.9\linewidth]{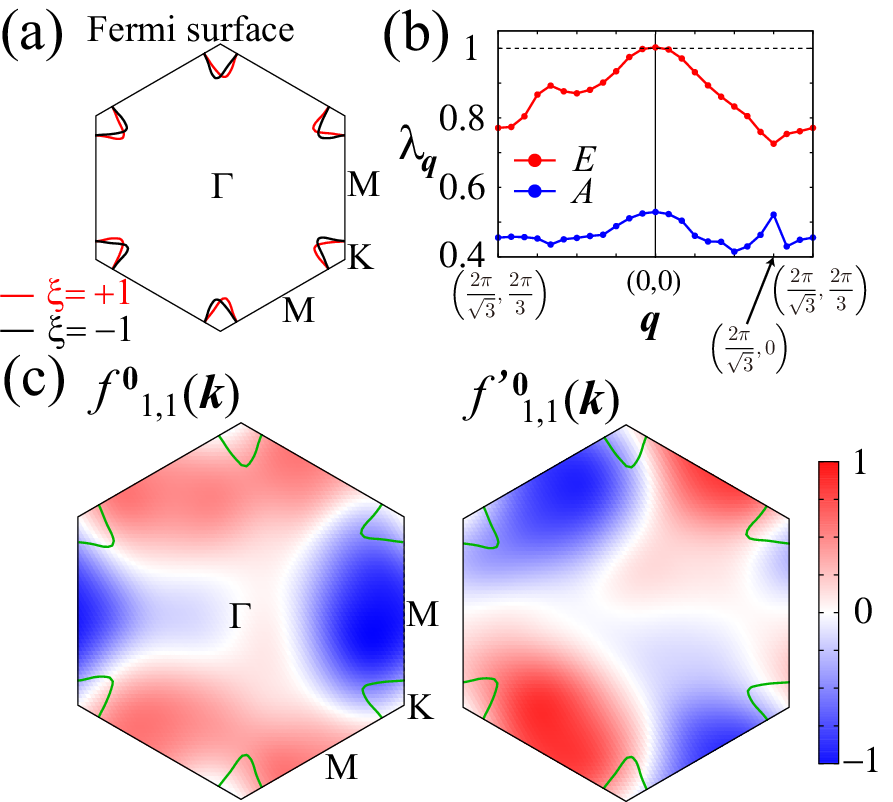}
 \caption{
 (a) FSs for $n=0.5$ in the original first-principles model, where red (black) lines denote the valley $\xi=+1$ $(-1)$. 
 (b) Obtained $\q$ dependences of $\lambda_{\q}$ for the $E$-symmetry \color{black}and
 $A$-symmetry \color{black}for $n=0.5$ in the
 original model.
(c) Doubly degenerate form factors $f_{1,1}^{\bm{0}}(\k)$ and $f'^{\bm{0}}_{1,1}(\k)$ in the Brillouin zone.
The green lines indicate FSs for $\xi=+1$. 
}
\label{fig:Original}
\end{figure}

In order to verify the robustness of the
nematic bond order, we investigate the original first-principles model\cite{S-Koshino}, by multiplying all
the hopping integrals by $50$ in order to fit
the bandwidth obtained by the STM
measurement \cite{S-nematic-TBG}. The FSs
and band structure are shown in Figs. \ref{fig:Original}(a) and \ref{fig:FS-n3.2}(a).
We stress that the band structure is similar to Fig. \ref{fig:FS}(d) in
the main text. In contrast, the FS structure is very different from
Fig. \ref{fig:FS}(c) in the main text, and the
VHS filling $n_{\rm VHS}=0.7$ is also very different from $n_{\rm
VHS}=2.07$ in the main text. Moreover, the positions of the VHS points are different. These differences mainly come from the
reduction of the imaginary intra-orbital hoppings.

Nonetheless of the large difference between two models, the nematic
state is also 
obtained in the original first-principles model when the filling is slightly lower than $n_{\rm VHS}$.
Figure \ref{fig:Original}(b) shows $\q$ dependences of the DW equation
eigenvalue $\lambda_{\q}$ for the $E$-symmetry and $A$-symmetry for
$n=0.5$ and $\a=0.91$ ($U=28.7$ meV) at $T=1.5$ meV.
The obtained $\lambda_{\q}$ for $E$-symmetry is dominant and has peak at $\q=\0$, which corresponds to the emergence
of the $\q=\0$ nematic order. The obtained doubly degenerate $E$-symmetry form factors
$f^{\bm{0}}_{1,1}(\k)$ and $f'^{\bm{0}}_{1,1}(\k)$ shown in
Fig. \ref{fig:Original}(c) are similar to those in
Fig. \ref{fig:Sigma}(b) in the main text and Fig. \ref{fig:deg}.
The real part Re$\tilde{f}_{1,1}(\r)$ gives the bond order, and
the imaginary part Im$\tilde{f}_{1,1}(\r)$ gives the spontaneous current.
Thus, the nematic state near the VHS filling is also identified as
the nematic bond order based on the original model.
The $E$-symmetry solution is doubly degenerate similarly to the results
in the model in main text.

In summary, although the value of $n_{\rm VHS}$ and the FS structure are very
different between the original first-principles model and the present model
in the main text, both models lead to essentially the same nematic bond
order solution.
Thus, $\q=\0$ nematic bond order is stably obtained near the VHS filling
irrespective of
huge difference in the FS structure. \color{black}This result is verified by the
analysis of the simple three VHS model in Fig \ref{fig:Sigma}(c) and SM
D 1. \color{black}

%%%%%%%%%%%%%%%%%%%%%%%%%%%%%%%%%
%%%%%%%%%%%%%%%%%%%%%%%%%%%%%%%%%
\color{black}
\subsection{E: Numerical study for $J\ne0$}
\begin{figure}[h]
\includegraphics[width=.9\linewidth]{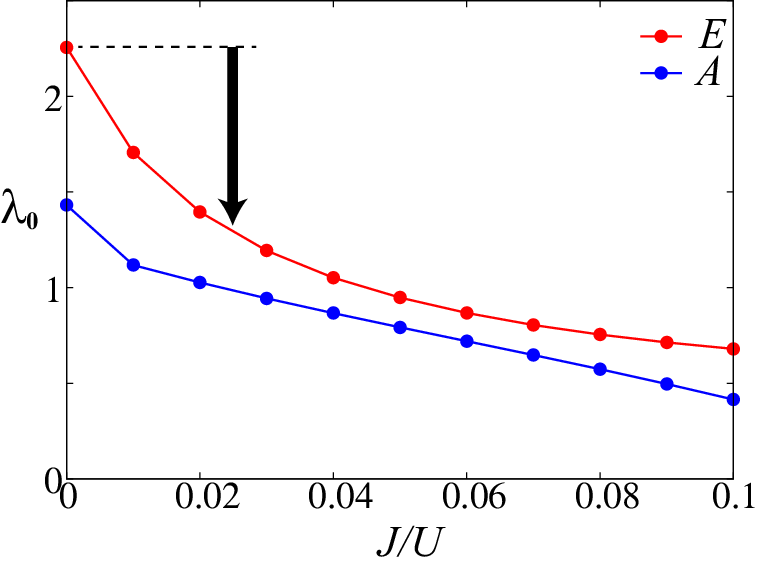}
 \caption{\color{black}
$J/U$ dependences of the DW eigenvalue
 $\lambda_{\bm{0}}$ for the $E$ and $A$ symmetries for $n=2.0$ and
 $\a_s=0.9$ at $T=1.5$meV. Black arrow shows strong suppression of
 $\lambda_{\bm{0}}$ for finite $J$.
}
\label{fig:J/U}
\end{figure}

Here, we show results for $J\ne0$ and $U=U'+2J$, which is satisfied in
the rotationally invariant systems. 
In the case $J\ne0$, $\alpha_s=\alpha_c$ is violated and $\alpha_s$
becomes larger than $\alpha_c$.
Figure \ref{fig:J/U} shows $J/U$ dependences of the DW eigenvalue
 $\lambda_{\bm{0}}$ for the $E$ and $A$ symmetries for $n=2.0$ by
 fixing $\a_s=0.90$. It is robust for $J/U<0.1$ that the $E$-symmetry nematic bond ordered state
  dominates over the $A$-symmetry ordered state.

However, the value of $\lambda_{\bm{0}}$ is
 strongly suppressed for $J/U\gtrsim 0.03$ as shown by arrow in
 Fig. \ref{fig:J/U}. This suppression is caused by the decreased valley
 fluctuation due to $\a_s>\a_c$ in finite $J$. Thus, $J=0$ and
 $U=U'$ are important features in MATBG to realize the nematic bond ordered
 state.
%%%%%%%%%%%%%%%%%%%%%%%%%%%%%%%%%
%%%%%%%%%%%%%%%%%%%%%%%%%%%%%%%%%
\color{black}

\subsection{F: $\q$ dependence of $\lambda_{\q}$ for $n=2.4$}
\begin{figure}[!htb]
\includegraphics[width=.8\linewidth]{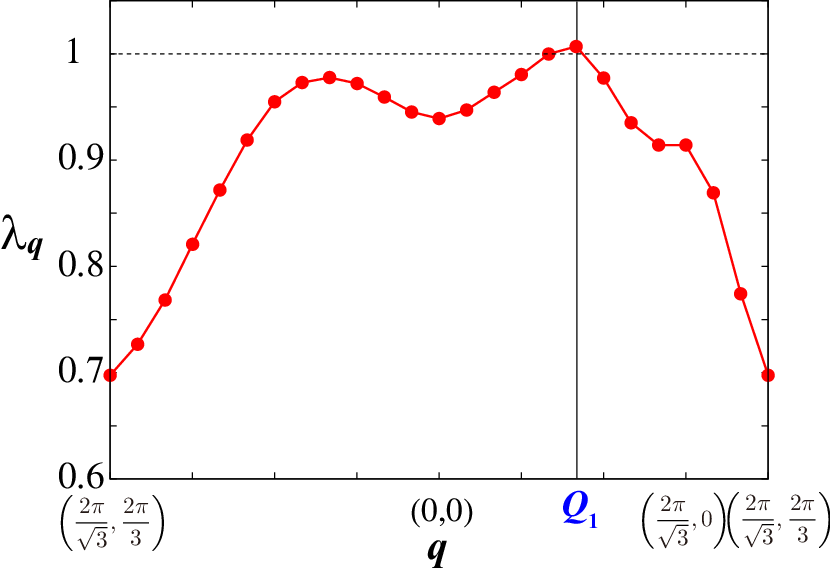}
\caption{
 $\q$ dependence of $\lambda_{\q}$ for $n=2.4$.
}
\label{fig:q-dep-n2.4}
\end{figure}

We discuss the $\q$ dependence of $\lambda_{\q}$ for $n=2.4$.
 Figure \ref{fig:q-dep-n2.4} shows $\lambda_{\q}$ obtained by the
 model in the main text. The $\lambda_{\q}$ has peak at $\q=\Q_1\sim
2\Q$ due to the quantum interference mechanism in Fig \ref{fig:FS}(a).
The obtained form factor $\hat{f}^{\q}(\k)$ violates
 time-reversal-symmetry. 
$\hat{f}^{\q}(\k)$ satisfies relations Re$f_{1,1}^{\q}(\k)$
 $={\rm Re}f_{2,2}^{\q}(\k)$ and Re$f_{1,1}^{\q}(\k)$ $=-{\rm
 Re}f_{3,3}^{-\q}(-\k)$. $\k$ dependence of $\hat{f}^{\q}(\k)$ is small
 as shown in Fig. \ref{fig:n3.2}(b). \color{black}

We confirm that $\lambda_{\Q_1}$ is enlarged by the Hartree term
in the MT terms and the AL type quantum
interference between the valley + spin fluctuations with
$\Q=\Q'$. 

%%%%%%%%%%%%%%%%%%%%%%%%%%%%%%%%
%%%%%%%%%%%%%%%%%%%%%%%%%%%%%%%%

\color{black}
\subsection{G: Smallness of spin-channel DW instabilities in MATBG}
In order to discuss the effect of the VCs for the spin-channel density waves,
we analyze the spin-channel DW equation
with spin-channel irreducible interaction,
${\hat I}^{s,\q=\bm{0}} \equiv{\hat I}^{\uparrow\uparrow,\q=\bm{0}}-{\hat I}^{\uparrow\downarrow,\q=\bm{0}}$,
where ${\hat I}^{\s\s',\q=\bm{0}}(k,k')= \delta \Phi/\delta {\hat G}^\s(k)\delta {\hat G}^{\s'}(k')$.
The DW equation is given as \cite{S-Kontani-sLC}
\begin{eqnarray}
  \eta_\q g^{\q}_{l,l'}(k)= \frac{T}{N}&&
 \!\!\sum_{k',m,m'} {K}^{s,\bm{q}}_{l,l';m,m'}(k,k')g^{\q}_{m,m'}(k'),
 \label{eqn:linearized2}\\ 
 K^{s,\bm{q}}_{l,l';m,m'}(k,k')=&&-\sum_{m_1,m_2}
 \!\!\!\!I^{s,\bm{q}}_{l,l';m_1,m_2}(k,k')G_{m_1,m}\left(k'+\frac{\bm{q}}{2}\right)\nonumber\\
 &&\times G_{m',m_2}(k'-\frac{\q}{2}),\label{eqn:SK}
\end{eqnarray}
where $\eta_{\q}$ and $\hat{g}^{\q}$ are the eigenvalue and the form factor of
the spin-channel DW equation, respectively.
The spin-channel $\hat{I}^{s,\q}$ is given as
\begin{eqnarray}
&& \!\!\!
I^{s,\bm{q}}_{l,l';m,m'}(k,k')=
\frac{1}{2} V^{s}_{l,m;l',m'}(k-k')-\frac{1}{2} V^{c}_{l,m;l',m'}(k-k')
\nonumber \\
&& 
+\frac{T}{N}\!\!\sum_{p,l_1,l_2,m_1,m_2}\!\!\!\!\!\!\!\!\!\!
 \left[V^{s}_{l,l_1;m,m_2}\left(p+\frac{\q}{2}\right)V^{s}_{m_1,l';l_2,m'}\left(-p+\frac{\q}{2}\right)\right.
\nonumber \\
&&\left.+\frac{1}{2}V^{s}_{l,l_1;m,m_2}\left(p+\frac{\q}{2}\right)V^{c}_{m_1,l';l_2,m'}\left(-p+\frac{\q}{2}\right)\right.
 \nonumber \\
&&
 \left.+\frac{1}{2}V^{c}_{l,l_1;m,m_2}\left(p+\frac{\q}{2}\right)V^{s}_{m_1,l';l_2,m'}\left(-p+\frac{\q}{2}\right)\right] 
\nonumber \\
&&
\times G_{l_1,m_1}(k-p)G_{l_2,m_2}(k'-p)
\nonumber \\
&&
+\frac{T}{N}\!\!\sum_{p,l_1,l_2,m_1,m_2}\!\!\!\!\!\!\!\!\!\!
 \left[ -V^{s}_{l,l_1;l_2,m'}\left(p+\frac{\q}{2}\right)V^{s}_{m_1,l';m,m_2}\left(-p+\frac{\q}{2}\right)\right.
 \nonumber \\
&&
+\frac{1}{2}V^{s}_{l,l_1;l_2,m'}\left(p+\frac{\q}{2}\right)V^{c}_{m_1,l';m,m_2}\left(-p+\frac{\q}{2}\right)
\nonumber \\
&&\left.+\frac{1}{2}V^{c}_{l,l_1;l_2,m'}\left(p+\frac{\q}{2}\right)V^{s}_{m_1,l';m,m_2}\left(-p+\frac{\q}{2}\right)\right]
\nonumber \\
&&
\times G_{l_1,m_1}(k-p)G_{l_2,m_2}(k'+p).
%-({\rm Double\;counting\;} [\hat{\Gamma}^{s(c)}]^2 \;{\rm terms})
\label{eqn:SDW} 
\end{eqnarray}

\begin{figure}[!htb]
\includegraphics[width=.99\linewidth]{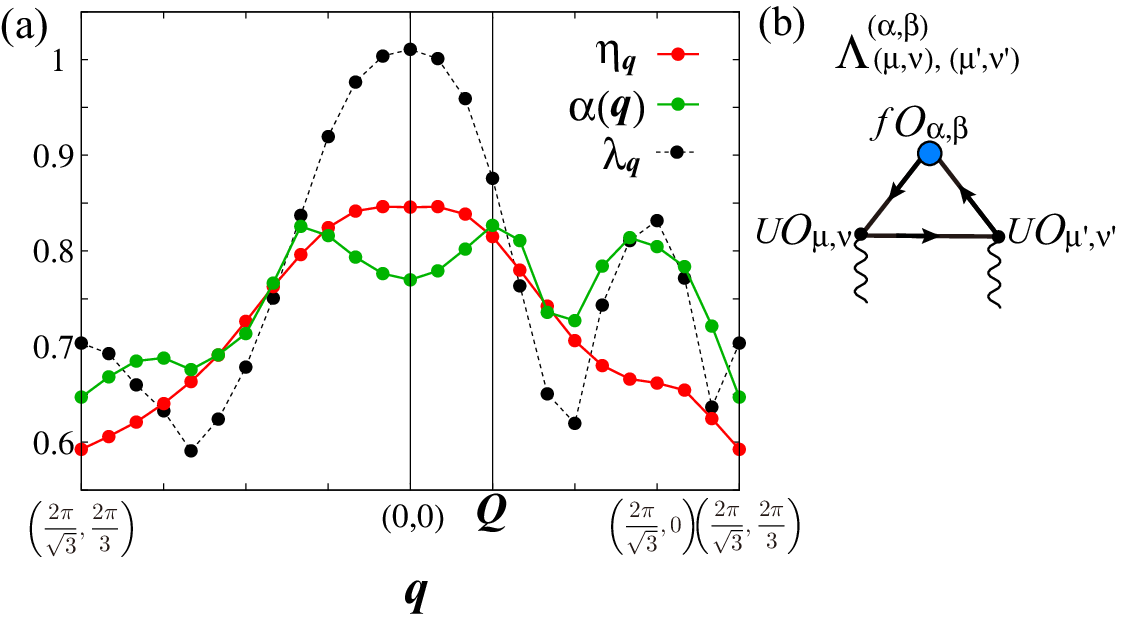}
\caption{\color{black}
 (a) $\q$ dependences of spin-channel eigenvalue $\eta_\q$ and the
 charge-channel one $\lambda_\q$ obtained by the DW equations. The RPA
 Stoner factor at $\q$, $\a(\q)$, is also shown. Model parameters are
 $n=2.0$ and $\alpha=0.83$.
(b) Three-point vertex in AL term, where $O_{\a,\b}$, $O_{\mu,\nu}$, and $O_{\mu',\nu'}$ are the spin-valley operators.
\color{black}
}
\label{fig:q-dep-spin}
\end{figure}

Figure \ref{fig:q-dep-spin}(a) displays $\q$ dependence of
the obtained spin-channel eigenvalue $\eta_{\q}$, together with
 $\alpha(\q)$ and the charge-channel eigenvalue $\lambda_{\q}$ for
$E$-symmetry shown in Fig. \ref{fig:Sigma}(a).
%The parameters are $n=2.0$ and $\alpha=0.83$.
Here, $\alpha(\q)$ is the RPA Stoner factor at fixed $\q$.
Note that $\alpha=\max_\q \alpha(\q)$.
We find the relation $\eta_{\q}\sim \a(\q)\sim 0.8$,
and the obtained spin-channel form factor $g^{\q={\bm0}}_{l,m}(k)$ is
nearly $k$-independent.
Because $\chi^s(\q) \propto 1/(1-\eta_{\q})$ is very similar to
$\chi^{{\rm RPA},s}(\q) \propto 1/(1-\a(\q))$, the VCs for the spin-channel density-waves are unimportant.
Therefore, the obtained nematic bond order is
robust even when both charge-channel and spin-channel VCs are taken into account.
%if the solution of spin-channel DW equation is used as the spin
% susceptibility $(\propto \frac{1}{1-\eta_{\q}})$.

Now, we explain the reason why the AL terms induce strong $O_{0,0}$-channel instability, based on the $O_{\mu,\nu}$-channel decomposition of
the Coulomb interaction in Eq. (\ref{eqn:HU})
by following Ref. \cite{S-Tazai-CeB6}.
The three-point vertex in the AL term $\Lambda^{(\a,\b)}_{(\mu,\nu),(\mu',\nu')}$ is shown in Fig. \ref{fig:q-dep-spin}(b),
where $f(k)O_{\a,\b}$ is the DW form factor
and $U O_{\mu,\nu}$ ($U O_{\mu',\nu'}$) represents the
decomposed Coulomb interaction in Eq. (\ref{eqn:HU}).
$O_{\mu,\nu}$ is converted to the $SU(4)$ valley + spin susceptibility
$\chi_{\mu,\nu}$ after taking the average.
%where $O_{\mu,\nu}$ is the endpoint of the $SU(4)$ valley + spin
%susceptibilities. Here, $(\mu,\nu) \ne(0,0)$. When $O_{\a,\b}$ order is
%induced by the AL terms, $\Lambda^{(\a,\b)}_{(\mu,\nu),(\mu',\nu')}$
%corresponds to the coupling constant of the interference between $O_{\mu,\nu}$ fluctuation and $O_{\mu',\nu'}$ fluctuation.
The three-point vertex represent the coupling constant between the DW form factor and the
$SU(4)$ valley + spin susceptibilities
in the interference mechanism. The relation
$\Lambda^{(\a,\b)}_{(\mu,\nu),(\mu',\nu')}\propto {\rm
Tr}[O_{\a,\b}O_{\mu,\nu}O_{\mu',\nu'}]$ holds because of the following relation in the Green function:
${\hat G}=G^a{\hat \s}_0{\hat \tau}_0$ ($a=AB$ or $BA$)
and $G^{AB}\approx G^{BA}$.
In the present mechanism, the interference between the same-channel fluctuations
[$(\mu,\nu)=(\mu',\nu') \ne (0,0)$] is particularly significant,
and $\Lambda^{(\a,\b)}_{(\mu,\nu),(\mu,\nu)}$
becomes nonzero only for $(\a,\b)=(0,0)$.
For this reason, the AL process induces the $O_{0,0}$-channel order selectively,
and the optimized form factor belongs to the $E$-symmetry
owning to the strong $\k$-dependence in the VCs,
as we explained in the main text. 

We note that the second largest charge-channel eigenvalue, $\lambda^{(2)}_{\q={\bm0}}$,
is equal to $\eta_{\q={\bm0}}$,
and the corresponding eigenstates are fifteen-fold degenerated.
As we discuss in SM B,
this interesting result is closely related to the approximate $SU(4)$ symmetry 
in MATBG \cite{S-Kang,S-Wang}.

%%%%%%%%%%%%%%%%%%%%%%%%%%%%%%%%
\color{black}

%The obtained solution of the spin-channel DW equation are twelvefold
%degenerate $A$-symmetry states.
%These solutions are represented by
%$f^{s,\q}_{l,l'}(k)\propto
%\sigma^\mu_{ss'}\tau^\nu_{\xi\xi'}$, where the orbital index $l(l')$
%includes the spin index
%$s(s')$ and the valley index $\xi(\xi')$. $\sigma^\mu$ and $\tau^\nu$
%are Pauli matrices for the spin and valley, respectively. The twelvefold form
%factors are given by $\mu=1\sim3$ and $\nu=0\sim3$. These solutions are
%also threefold degenerate with the charge-channel $A$-symmetry solutions
%$f^{\q}_{l,l'}(k)\propto \sigma^0_{ss'}\tau^\nu_{\xi\xi'}$ for $\nu=1\sim 3$.
%These fifteenfold degenerate $A$-symmetry states are mainly caused by
%the MT term, and related to the $SU(4)$ symmetry in MATBG\cite{S-Kang,S-Wang}.
%\color{black}

%%%%%%%%%%%%%%%%%%%%%%%%%%%%%%%%%%%%%%
\color{black}
\subsection{H: Effects of off-site Coulomb interaction on the nematic order}
% in Kang-Vafek model}

In the main text, we studied the MATBG Hubbard model with 
on-site Coulomb interaction $U$.
However, off-site Coulomb interactions are not small
because of the large size of the Wannier function in MATBG;
its size exceeds the AB-BA distance in Fig. \ref{fig:FS} (b) \cite{S-Koshino,S-Kang}.
The Coulomb interaction term is given as
\begin{eqnarray}
H' = U\sum_{i}n_{i\uparrow}n_{i\downarrow}
+\frac12 \sum_{i\ne j}V_{i,j}n_{i}n_{j},
\end{eqnarray}
where $n_{i\s}$ is the $\s$-spin electron number operator at site $i$,
$n_{i}=n_{i\uparrow}+n_{i\downarrow}$,
and $V_{i,j}$ is the off-site Coulomb interaction.
(We drop the inter-site density terms.) 
Authors in Ref. \cite{S-Kang} derived the 
Coulomb interaction by considering the screening due to the metallic gate.
Then, approximately, the ratio between the nearest, the next nearest, and the third nearest
Coulomb potential is $V_1:V_2:V_3=2:1:1$, 
and the ratio $r \equiv V_1/U$ is $2/3$.

In the main text, we studied the case of $r=0$.
Now, we discuss the effect of the 
off-site Coulomb interaction ($r\ne0$) on the nematicity.
Here, we study the case of $r\ll 2/3$ because 
the effective $r$ will be reduced by
the Thomas-Fermi screening due to the 
conduction electrons of MATBG.

%Now, in order to discuss the role of $V_n$ ($n=1,2,3$),
Hereafter,
we consider $r$ as a parameter
and fix the ratio $V_1:V_2:V_3=2:1:1$ for simplicity.
The bare interaction by $V_n$ is expressed as
\begin{eqnarray}
I^{V,\q}_{ll',mm'}(k,k')&=&-2V_{l,m}(\q)\delta_{l,l'}\delta_{m,m'} \nonumber\\
&& + V_{l,l'}(\k-\k')\delta_{l,m}\delta_{l',m'},
\nonumber\\
\label{eqn:IV}
\end{eqnarray}
where $V_{l,m}(\k)$ is the Fourier transformation of $V_{i,j}$,
and $l,m=1\sim4$.
The first and the second terms in Eq. (\ref{eqn:IV})
correspond to the Hartree and the Fock terms, respectively.

\begin{figure}[!htb]
\includegraphics[width=.6\linewidth]{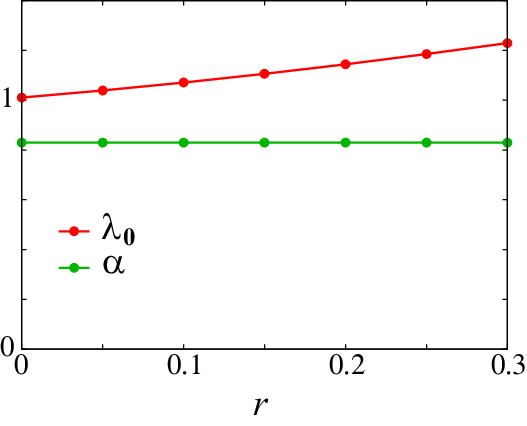}
\caption{
Obtained charge-channel eigenvalue $\lambda_{\q={\bm0}}$ 
and Stoner factor $\a$ as functions of $r=V_1/U$. $\lambda_{\bm{0}}$ linearly
 increases with $r$, while $\a$ is independent of $r$.
}
\label{fig:VU}
\end{figure}

Here, we introduce ${\hat I}^{V,\q}$ in Eq. (\ref{eqn:IV})
into the irreducible interaction ${\hat I}^{c,\q}$ in Eq. (\ref{eqn:S-K})
to discuss the effect of $V_n$ on the nematicity.
(Unfortunately, serious diagrammatic calculation 
of the MT and AL terms in the $U+\{V_n\}$ Hubbard model
is very difficult.)
Figure \ref{fig:VU} exhibits the eigenvalue 
of charge-channel DW equaiton
with the irreducible interaction
${\hat I}^{c,\q}(k,k')+{\hat I}^{V,\q}(k,k')$.
%The former and the latter terms are given by Eqs. (\ref{eqn:S-K})
%and (\ref{eqn:IV}), respectively.
We see that the nematic order eigenvalue at $\q={\bm0}$
linearly increases with $r \equiv V_1/U$.
Thus, the nematic order due to the AL process
is stabilized by finite $V_n$,
due to the Fock term in Eq. (\ref{eqn:IV}).
(In contrast, the Stoner factor $\a$ is independent of $V$ for $r<0.3$.)

%%%%%%%%%%%%%%%%%%%%%%%%%%%%%%%%
%%%%%%%%%%%%%%%%%%%%%%%%%%%%%%%%%%%%%%

Thus, we conclude that the driving force of the ferro-nematic order in MATBG 
is the valley + spin fluctuation interference mechanism,
and finite off-site Coulomb potential will stabilize the nematic order.
We stress that the Hartree-Fock term alone in Eq. (\ref{eqn:IV}) yields the density-wave order with $\q \ne \bm{0}$.
Note that $\a_c=1$ is satisfied when $r=0.75$ in the RPA
Therefore, the present $SU(4)$ interference mechanism is
essential for the $\q = \bm{0}$ nematic order. 
\color{black}

%%%%%%%%%%%%%%%%%%%%%%%%
%references
%%%%%%%%%%%%%%%%%%%%%%%%

\end{document}